\documentclass[]{aa}  
\bibpunct{(}{)}{;}{a}{}{,}
\usepackage{graphicx}
\usepackage{txfonts}
\usepackage{enumitem}
\usepackage{epstopdf}
\usepackage[percent]{overpic}
\usepackage{comment}
\usepackage{color}
\usepackage{url}
\usepackage[rightcaption]{sidecap}
\usepackage{breqn}
\usepackage{dsfont}
\usepackage{natbib}
\usepackage{footmisc}
\usepackage[colorlinks=true, allcolors=blue,citecolor=blue,linkcolor=blue]{hyperref}
\usepackage{xcolor}
\newcommand{\ca}{\mbox{Ca\,{\textsc{ii}}~K\,}}

\begin{document} 
\sloppy
\title{Re-evaluating solar irradiance reconstructions: No evidence for large secular trends}
\titlerunning{Re-evaluating solar irradiance reconstructions: No evidence for large secular trends}  
\authorrunning{Chatzistergos et al.}

\author{Theodosios Chatzistergos\inst{1}, Natalie A. Krivova\inst{1}, Mathew J. Owens\inst{2}, Tatiana A. Egorova
\inst{3}} 
\institute{Max Planck Institute for Solar System Research, Justus-von-Liebig-Weg 3,	37077 G\"{o}ttingen, Germany \and Department of Meteorology, University of Reading, Earley Gate, Reading, RG6 6BB, UK \and Physikalisch-Meteorologisches Observatorium Davos/World Radiation Center (PMOD/WRC), Davos, Switzerland} 
\offprints{Theodosios Chatzistergos  \email{chatzistergos@mps.mpg.de}}


\abstract
{Constraining long-term variations in total solar irradiance (TSI) is important for assessing the solar influence on Earth's climate.
Most current irradiance reconstructions suggest only a modest secular increase in TSI between the 1700 and 1986 activity minima ($<1$~W\,m$^{-2}$), while two models suggest substantially larger changes of 2.1--5.9~W\,m$^{-2}$.
These are the Code for the High spectral ResolutiOn recoNstructiOn of Solar irradiance (CHRONOS) and the model by \citet[][PEA24]{penza_reconstruction_2024}.
Although the two models differ substantially in their architecture, both use cosmogenic isotope records combined with neutron monitor data to describe the long-term irradiance variability. 
However, cross-calibrating modulation potential reconstructions from cosmogenic isotopes against neutron monitor data remains highly uncertain.}
{We reassess the origin and the magnitude of the large secular trends in the CHRONOS and PEA24 models.} 
{We update the CHRONOS and PEA24 models using recent heliospheric modulation potential and open solar flux reconstructions based on geomagnetic data and neutron monitor measurements, which provide a more reliable connection between cosmogenic isotope and neutron monitor records.
We further apply a more robust smoothing methodology for the long-term series.
These allow the secular component of the reconstructions to be extended consistently to the satellite era and compare the resulting TSI reconstructions with direct TSI measurements.
}
{We find that the original CHRONOS and PEA24 reconstructions substantially overestimated the secular variability in irradiance.
When constrained by direct TSI measurements, CHRONOS yields a TSI increase of about 0.1--0.44~W\,m$^{-2}$ between the 1700 and 1986 minima, while PEA24 returns  about 0.22~W~m$^{-2}$.
Our analysis indicates that the previously inferred large secular trends arose primarily from improper linking cosmogenic isotope and neutron monitor records, together with issues in the adopted smoothing approach. 
}
{The updated reconstructions presented here point toward a relatively modest secular increase in TSI since the Maunder Minimum, likely below 1~W\,m$^{-2}$, consistent with the majority of current irradiance models.}
   \maketitle

\section{Introduction}
\label{sec:introduction}
The Sun is the main external energy source to the Earth system \citep[e.g.][]{kren_where_2017}, making solar variability an important driver of climate variability \citep{haigh_sun_2007,gray_solar_2010,solanki_solar_2013-1,masson-delmotte_climate_2021}.
Direct measurements of the solar radiative energy output, quantified through solar irradiance at the top of Earth's atmosphere and normalised to the mean Sun--Earth distance, started only in the late 1970s \citep[see reviews by][]{ermolli_recent_2013,kopp_solar_2025}.
Because this direct observational period is short compared to climate-relevant timescales, reconstructing past irradiance variations is necessary for understanding the solar impact on Earth's climate.

On timescales from days to millennia, irradiance variations are generally attributed to the evolution of the solar surface magnetic field \citep{krivova_reconstruction_2003,shapiro_nature_2017,yeo_solar_2017}.
Irradiance reconstructions therefore rely on proxies describing the competing effects of bright faculae and dark sunspots.
While sunspot observations extend back to the early 17th century, direct facular measurements are much more limited in time \citep[e.g.][]{chatzistergos_full-disc_2022,chatzistergos_understanding_2024,clette_recalibration_2023}.
As a result, estimates of long-term irradiance variability depend strongly on how facular changes are represented prior to the modern observational era.

Various irradiance models have been developed with varying complexity and performance \citep[see, e.g., reviews by][]{domingo_solar_2009,ermolli_recent_2013,chatzistergos_long-term_2023}.
Most existing reconstructions indicate a relatively modest secular increase in total solar irradiance (TSI) between the Maunder Minimum and the present \citep[$<$1~W\,m$^{-2}$; see review by][]{chatzistergos_long-term_2023}, although some models suggest changes exceeding 5~W\,m$^{-2}$ \citep[e.g.][]{shapiro_new_2011,egorova_revised_2018}.

Reconstructions extending to multi-centennial and millennial timescales commonly rely on cosmogenic isotope records to constrain long-term variability \citep[e.g.][]{lean_estimating_2018,egorova_revised_2018,wu_solar_2018-2,penza_reconstruction_2024,temaj_reconstruction_2026}.
The most widely used cosmogenic isotopes are $^{14}$C and $^{10}$Be \citep[see e.g.][]{beer_cosmogenic_2012,usoskin_history_2023}. 
These isotopes are produced primarily through interactions between galactic cosmic rays and target nuclei in the Earth’s atmosphere.
Their production rates therefore depend on the cosmic-ray flux, which is modulated by solar magnetic activity through heliospheric shielding, as well as by the geomagnetic field.
After production, $^{14}$C is rapidly oxidised to carbon dioxide and enters the global carbon cycle \citep[e.g.][]{beer_cosmogenic_2012,usoskin_history_2023}, enabling its measurement in tree rings. 
Its reliability is significantly reduced since about 1880 due to anthropogenic effects, most notably large-scale fossil fuel combustion \citep{suess_radiocarbon_1955} and nuclear bomb tests in the late 1940s.
$^{10}$Be attaches to atmospheric aerosols before being deposited and preserved in natural archives such as ice cores \citep[e.g.][]{beer_cosmogenic_2012,baroni_persistent_2019}.
Its deposition is influenced by atmospheric transport, climate variability, and volcanic eruptions \citep[e.g.][]{baroni_persistent_2019}. 
Extracting the solar signal from the radionuclide records therefore requires disentangling solar, geomagnetic, and climatic effects.

Cosmogenic isotope records are widely used to derive quantities such as the solar modulation potential, open solar flux (OSF), or sunspot numbers, which are then employed as inputs for irradiance reconstructions. 
Because cosmogenic isotope records become increasingly uncertain in the modern era, neutron monitor data have often been used to extend these data to the present.
However, neutron monitor measurements started only in 1951, and thus linking them to cosmogenic isotope data carries significant uncertainty \citep[see e.g.][]{belov_large_2000,usoskin_history_2023}.
Sometimes the neutron monitor record is extended back to 1936 using measurements from Forbush ground-based ionisation chambers \citep{mccracken_long-term_2007}. 
However, these data carry large uncertainties \citep[see e.g.][]{usoskin_solar_2011} and may introduce further biases when connected with cosmogenic isotope records.

These issues are particularly important for the two contemporary irradiance models that produce secular TSI changes exceeding 2~W\,m$^{-2}$ between the 1700 and 1986 activity minima. 
These are the CHRONOS model \citep[Code for the High spectral ResolutiOn recoNstructiOn of Solar irradiance;][]{shapiro_new_2011,egorova_revised_2018} and the model by \citet[][PEA24, hereafter]{penza_total_2022,penza_reconstruction_2024}.
CHRONOS yields a TSI increase of approximately 3.5--5.5~W\,m$^{-2}$ between the 1700 and 1986 minima, while PEA24 suggests a change of about 2.2~W\,m$^{-2}$.
Although the architecture of the models is quite different, with CHRONOS being semi-empirical and PEA24 empirical, both employ modulation potentials derived from cosmogenic isotopes to prescribe the secular component of irradiance variability.

Here we reassess the secular variability in the CHRONOS and PEA24 models using the recent modulation potential reconstruction by \citet{owens_geomagnetic_2024} and the OSF reconstruction by \citet{lockwood_reconstruction_2024}, which connect the cosmogenic isotope records with neutron monitor data more accurately than previously. 
In addition, we implement a more robust smoothing methodology for the time series.
Together, these improvements allow a more reliable assessment of the plausible range of secular TSI variability.

The paper is structured as follows.
Section~\ref{sec:data} describes the datasets used in this study. 
Sections~\ref{sec:chronos} and \ref{sec:penza} describe the CHRONOS and PEA24 models, respectively, together with our updated TSI reconstructions and comparisons with direct TSI measurements and other irradiance models. 
We discuss our results and their implications in Sect.~\ref{sec:discussion} and summarise our conclusions in Sect.~\ref{sec:summary}.

\section{Data}
\label{sec:data}

\subsection{Sunspot and facular data}
\label{sec:sunspotnumber}
To allow a consistent comparison with the original CHRONOS model, we employed exactly the same sunspot number record as used by \citet{egorova_revised_2018}.
It includes data from three datasets: the international sunspot number version 2 \citep[ISNv2;][]{clette_recalibration_2023}, the group number by \citet[][CEA17]{chatzistergos_new_2017} and the sunspot number series by \citet[][LEA14]{lockwood_centennial_2014}. 
ISNv2 is maintained by the Sunspot Index and Long-term Solar Observations (SILSO)\footnote{\url{https://www.sidc.be/SILSO/datafiles}} and provides daily, monthly, and annual values back to 1818, 1749, and 1700, respectively.
CEA17 provides daily values back to 1739, but data before 1749 are sparse and therefore not considered here.
LEA14 covers the period back to 1610 with annual values.
Following \citet{egorova_revised_2018}, we used ISNv2 since 1900, extended back to 1749 with CEA17 scaled to ISNv2, and further back to 1612 with LEA14 scaled to CEA17. 
We stress that using a different sunspot record, or different scaling between the series, would not affect our results about the secular trend in TSI, since by the design of the CHRONOS model this choice only influences the solar cycle irradiance variations and does not affect the secular trend.

To allow a consistent comparison with the original PEA24 model, we use the sunspot areas by \citet{mandal_sunspot_2020} and the plage areas by \citet{chatzistergos_analysis_2020}.
The plage area series of \citet{chatzistergos_analysis_2020} is a composite constructed from 38 \ca archives, spanning the period 1892--2024.
However, due to the sparsity of data prior to 1903, we follow \citet{penza_reconstruction_2024} in excluding the earlier period.
\citet{mandal_sunspot_2020} provide a composite sunspot area series, combining the Royal Greenwich Observatory (1874--1976), Debrecen (1976--2018), and Kislovodsk (1977--2024) records. 
There are 329 days between 28 June 2018 and 7 September 2025 without records in the \citet{mandal_sunspot_2020} composite, mostly due to incorrectly classified spotless days in the source archives over this period.
To fill these gaps and include values beyond 7 September 2025 (the last day included in the \citealt{mandal_sunspot_2020} series) we additionally use sunspot areas derived from Helioseismic and Magnetic Imager (HMI) continuum observations onboard the Solar Dynamics Observatory \citep[SDO;][]{scherrer_helioseismic_2012}, as processed by \citet{chatzistergos_revisiting_2025}.
Both plage and sunspot areas used here are projected (i.e., not corrected for foreshortening) and are expressed as fractions of the visible solar surface.
We also use the group sunspot number by \citet{hoyt_group_1998} covering the period 1610--1995.
We use this series because it has daily values back to the Maunder minimum, but we note on one hand caveats with it \citep[e.g.][]{clette_recalibration_2023,chatzistergos_assessment_2025} and on the other hand that this has no effect on the secular trend in our reconstruction.

\begin{figure*}[]
  {	\centering
	\begin{overpic}[width=0.95\linewidth,trim={0 0.9cm 0cm 0.0cm},clip]{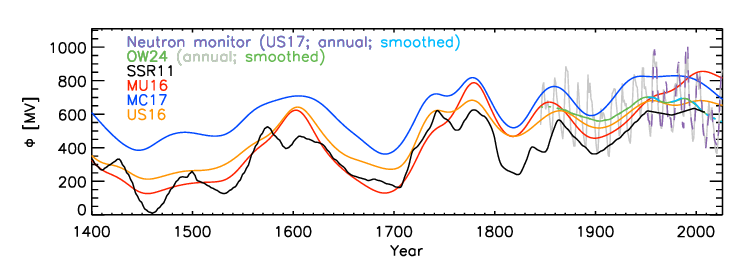}
	    \put (14.5,23) {a)}     \end{overpic}
	\begin{overpic}[width=0.95\linewidth,trim={0 0.9cm 0cm 0.4cm},clip]{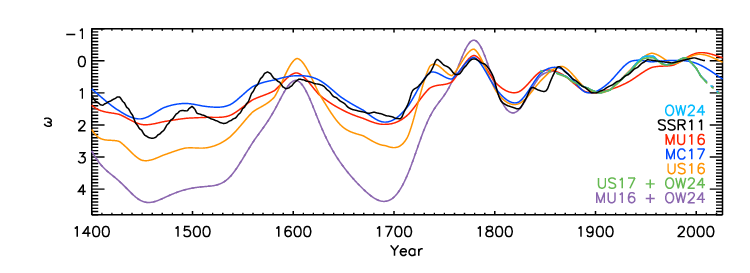} \put (14.5,23) {b)}     \end{overpic}
	\begin{overpic}[width=0.95\linewidth,trim={0 0.9cm 0cm 0.4cm},clip]{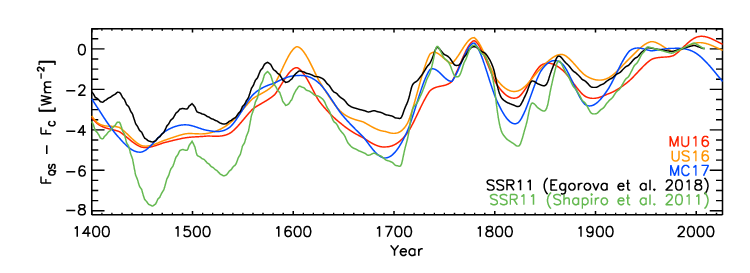} \put (14.5,23) {c)}     \end{overpic}
	\begin{overpic}[width=0.95\linewidth,trim={0 0.cm 0cm 0.4cm},clip]{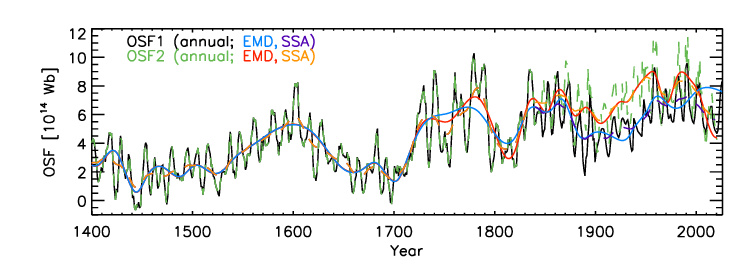} \put (14.5,30.2) {d)}     \end{overpic}
 	  \caption{
      \textit{Panel a):} Solar modulation potential series used in this study. 
      \textit{Panel b):} Normalised modulation potential: Parameter $\omega$ (Eq. \ref{eq:omega}); see Sect. \ref{sec:model} for a complete definition.
      \textit{Panel c):} Estimated secular component of the quiet-Sun irradiance, $F_{\mathrm{QS}}\left(t\right)-F_C$, required for our CHRONOS reconstruction to reproduce the published TSI reconstructions of \citet{egorova_revised_2018} for the corresponding  modulation potential, as well as that of \citet[][green]{shapiro_new_2011}.
      \textit{Panel d):} Open solar flux, OSF, series used in this study (see Sect.~\ref{sec:modpotential}. The empirical mode decomposition (EMD) is described in Sect.~\ref{sec:penzaseculartrend}, while the singular spectrum analysis (SSA) filter in Sect. \ref{sec:modpotential}.}}
      \label{fig:modpotentials}
\end{figure*}

\subsection{Modulation potentials and open solar flux series}
\label{sec:modpotential}

We use seven modulation potentials:
\begin{enumerate}
	\item \citet[][US16; covering $-7439$--2021]{usoskin_solar_2016}
	\item \citet[][MU16; covering 19--2021]{muscheler_revised_2016}
	\item McCracken and Beer \citep[2017; unpublished record provided to][MC17; covering $-7439$--2021]{egorova_revised_2018}
	\item \citet[][SSR11; covering $-500$--2008]{shapiro_new_2011}. This series combines the $^{10}$Be-based reconstructions by \citet{mccracken_geomagnetic_2004} and \citet{vonmoos_large_2006}.
	\item \citet[][covering 1000--2001]{muscheler_solar_2007}; an earlier version of MU16 combined with neutron monitor \citep{masarik_simulation_1999} and ionisation chamber data since 1937.
	\item Neutron monitor-based by \citet[][NMU17; covering 1951--2025]{usoskin_heliospheric_2017}\footnote{\url{https://cosmicrays.oulu.fi/phi/Phi_Table_2017.txt}}.
	\item Geomagnetic data-based by \citet[][OW24; covering 1845--2020]{owens_geomagnetic_2024}
\end{enumerate}

The first four series are those used by \citet{egorova_revised_2018}. 
For the period after 1950, these were extended using the modulation potential derived from neutron monitor data by \citet{usoskin_solar_2005}, an earlier version of NMU17. 
\citet{shapiro_new_2011} and \citet{egorova_revised_2018} incorporated these neutron monitor data as 22-year averages centred on 1977, 1999, and 2021.
We note, that this required an estimate of the modulation potential in the future, and that the choice of the central year also influences the result (see Appendix~\ref{appendix:smoothing}), both introducing uncertainty.
We digitised the SSR11 series from Figure~2 of \citet{shapiro_new_2011}, while the US16, MU16, and MC17 records were provided by the authors of \citet{egorova_revised_2018} as 22-year averages after they combined them with the neutron monitor data.
We interpolated the US16, MU16, and MC17 22-year averages to annual cadence.
Figure \ref{fig:modpotentials}a) shows these modulation potential series highlighting the differences between them.
Note that the modulation potential is not a physical quantity, but rather a formal parameter of the force-field approximation for cosmic-ray modulation, and its value depends on the assumed unmodulated cosmic-ray spectrum, the so-called local interstellar spectrum (LIS). 
Accordingly, different series, such as those used here, must first be reduced to a common reference condition before they can be directly compared or combined \citep{usoskin_heliospheric_2005,asvestari_neutron_2017}. 

OW24 is a reconstruction of the solar modulation potential covering the period 1845--2020 based on geomagnetic data.
This builds on the model of \citet{asvestari_empirical_2016}, in which
\begin{equation}
\phi = \phi_0F_s^{n}\left(1+A\sin{\alpha}(1+B p)\right),
\end{equation}
where $F_s$, $\alpha$,  and $p$ are the open solar flux (in units of 10$^{15}$~Wb), the heliospheric current sheet tilt angle, and the polarity ($\pm1$), respectively.
The free parameters $\phi_0$, $n$, A, and B were determined by calibration to a reference series based on neutron monitor data.
$F_s$ was reconstructed from the geomagnetic aa-index \citep{lockwood_reconstruction_2024}, while $\alpha$ and $p$ were inferred from the solar cycle phase.
This new dataset substantially reduces the uncertainty in linking cosmogenic isotope records with neutron monitor data.
As shown in Fig.~\ref{fig:modpotentials}a), the series closely matches NMU17, which we use to extend it to 2025.
We use OW24 smoothed with a 22-year low-pass singular spectrum analysis (SSA) filter.
Both the annual and smoothed series are shown in Fig.~\ref{fig:modpotentials}a) (see Appendix \ref{appendix:smoothing} for a discussion on the choice of the smoothing approach).

Finally, we additionally construct two combined series.
In the first, we use the OW24 reconstruction up to 1951 and the NMU17 series \citep{usoskin_heliospheric_2017} from 1952 onwards.
This avoids potential uncertainties in the OW24 series over the period used for comparisons with direct TSI measurements, while extending the record to present.
In the second, we combine the OW24 modulation potential all the way back to 1845 with the MU16 series for earlier times.
This dataset minimises the potential impact of the Suess effect by using the OW24 series over the period when such an influence may be present in cosmogenic isotope-based reconstructions.
It retains isotope-based values for earlier periods when this effect is not present. 
Since MU16 reaches the lowest values during the Maunder minimum among the series considered here, this combination yields the extreme case with a relatively deep minimum compared to the early 20th century.

We use two open solar flux (OSF) series. 
The first one is by \citet{usoskin_solar_2021} based on annual $^{14}$C measurements by \citet{brehm_eleven-year_2021} covering the period 971--1899.
Following \citet{penza_reconstruction_2024} we extended it to present by converting modulation potential series to open solar flux based on the equation:
\begin{equation}
    \mathrm{OSF}=(0.008\pm0.002) \phi + (1.4\pm0.9) 10^{14} \mathrm{Wb}.
    \label{eq:phitoosf}
\end{equation}
In particular, \citet{penza_reconstruction_2024} used the open solar flux by \citet{usoskin_solar_2021} up to 1899, the scaled modulation potential by \citet{muscheler_solar_2007} between 1900 and 1950, and the NMU17 modulation potential from the period after 1950.
To simplify the following discussion, we call this compilation as OSF1. 

The second OSF series we use is the one by \citet{lockwood_reconstruction_2024} based on geomagnetic data covering the period 1845--2020.
Here, we combine the OSF reconstruction of \citet{usoskin_solar_2021} up to 1844 with that of \citet{lockwood_reconstruction_2024} for subsequent periods, which we will refer to as OSF2.
Both OSF1 and OSF2 are shown in Fig.~\ref{fig:modpotentials}d).

\subsection{TSI}
\label{sec:TSI}

The TSI datasets used in this study include both composites of direct measurements and model-based reconstructions.
To estimate the uncertainty in direct measurements of TSI we consider the range covered by the existing composites.
The composites considered here are those by \citet[][CPMDF, hereafter; we use two different versions, the one from 20 January 2023 and the most recent one available from 19 November 2025. If no date is mentioned then we refer to the latest version. The series extend back to 1980]{montillet_data_2022}, ACRIM \citep[Active Cavity Radiometer Irradiance Monitor, being the instrument used as the reference;][covering 1978--2013]{willson_total_1997,willson_composite_2003}, ROB \citep[for Royal Observatory of Belgium;][1992--2026]{dewitte_total_2016}, \citet[][DdWK17, hereafter; 1978--2025]{dudok_de_wit_methodology_2017}, and C3S\footnote{\url{https://confluence.ecmwf.int/pages/viewpage.action?pageId=304239360}} (for Copernicus Climate Change Service; 1979--2026).
We note, that while most composites are regularly updated, ACRIM has not received an official update since 2013. 

For completeness, we also consider all 21 series compiled by \citet{connolly_multiple_2024}.
These include earlier versions of the \mbox{CPMDF}, C3S, DdWK17, ACRIM and PMOD \citep[Physikalisch-Meteorologisches Observatorium Davos;][]{frohlich_solar_2006}, as well as extensions of DdWK17, ACRIM, and PMOD using additional sources of TSI measurements.
However, these extensions were not constructed using the original methodologies of the respective teams. 
Furthermore, the ensemble is strongly weighted towards ACRIM-based series, with 16 out of 21 relying on ACRIM, while only two are based on DdWK17 and one each on the remaining composites.
In addition, some of the ACRIM series included are outdated and have been superseded by later revisions, including a 2011 version that exhibits a larger ACRIM-gap than the subsequently updated 2013 version
\citep[for discussions about the ACRIM-gap see][]{krivova_acrim-gap_2009,amdur_negative_2025,chatzistergos_revisiting_2025}.
As a result, the \citet{connolly_multiple_2024} set does not represent a balanced or internally consistent ensemble of TSI composites and places disproportionate emphasis on the ACRIM-gap change. 
Nevertheless, we include all 21 series to provide a deliberately conservative upper bound on the uncertainty range.
Our goal here is not to assess individual composites, but to bracket the plausible range of TSI variability.
All series are offset to match their annual mean values over a selected activity minimum (either 2009 or 1986). 
For each year, we then determined the mean, minimum, and maximum TSI values across all the selected composites, yielding lower and upper envelopes over 1978--2026.
Two such estimates were constructed.
The first includes only the \mbox{CPMDF}, DdWK17, ROB, and C3S, which we regard as the most reliable composites.
The second additionally includes ACRIM and all 21 series from \citet{connolly_multiple_2024}, providing an extremely conservative uncertainty range.
For both estimates, we also considered the 1-$\sigma$ uncertainty interval reported by DdWK17, although we note that it is substantially smaller than the spread among the 21 series.

Finally, we also compared our results with earlier CHRONOS-based reconstructions by \citet[][covering 850--2009]{shapiro_new_2011} and \citet[][1610--2015]{egorova_revised_2018}, as well as the  reconstructions by \citet[][971--2022]{penza_reconstruction_2024}, SATIRE-T \citep[Spectral And Total Irradiance REconstruction, `T' referring to the telescopic era;][1611--2025]{temaj_solar_2026}, \citet[][1700--2020]{dewitte_centennial_2022}, NNL \citep[for National oceanic and atmospheric administration~-- National aeronautics and space administration~-- Laboratory for atmospheric and space physics;][1874--2024]{deland_spectral_2026}, \citet[][695--1982]{delaygue_antarctic_2011}, and the \citet{hoyt_discussion_1993} model as recently updated by \citet[][1706--2010]{chatzistergos_discussion_2024}.

\section{Reassessing CHRONOS}
\label{sec:chronos}

In this section, we first reformulate the CHRONOS model in a way that isolates the contribution of the quiet Sun to long-term variability and allows the secular component of the reconstruction to be examined explicitly. 
The reformulation presented below is mathematically equivalent to the original CHRONOS formulation, with only minor technical modifications required for reproducibility.

\subsection{Model description}
\label{sec:model}
CHRONOS is a semiempirical solar irradiance model initially introduced by \citet{shapiro_new_2011} and later updated by \citet{egorova_revised_2018}.
In reproducing the model, we follow the published CHRONOS methodology as closely as possible.
However, some of the original component spectra used in the published reconstructions are no longer publicly available and could not be fully recovered.
We therefore use similar corresponding atmosphere models available in the literature.
These substitutions have only a minor effect on TSI and do not affect the conclusions regarding the secular variability discussed below.

In CHRONOS, the spectral solar irradiance, $F(\lambda,t)$, is computed as:
\begin{align}
\label{eq:rec}
    F(\lambda,t) = F_{\mathrm{QS}}\left(\lambda,t\right)  + \alpha_{\mathrm{f}}(t)(F_{\mathrm{f}} \left(\lambda\right)-F_{\mathrm{QS}}\left(\lambda,t\right))    \nonumber \\
    +\alpha_{\mathrm{su}}(t)(F_{\mathrm{su}}\left(\lambda\right)-F_{\mathrm{QS}}\left(\lambda,t\right))    \\
    +\alpha_{\mathrm{sp}}(t)(F_{\mathrm{sp}}\left(\lambda\right)-F_{\mathrm{QS}}\left(\lambda,t\right)), \nonumber
\end{align}
where $\alpha$ are the filling factors (see Sect.~\ref{sec:ff}), and the indices QS, f, su, and sp refer to quiet Sun, faculae, umbra, and penumbra, respectively.
The fluxes, $F$, of faculae, umbra, and penumbra are assumed to be time-invariant, as in other semi-empirical and physics-based irradiance models.
As noted above, some of the original model atmospheres and the corresponding intensity spectra used in CHRONOS are unavailable.
For the present reconstruction, we therefore represent sunspot umbrae and penumbrae using the \citet{kurucz_new_1993,kurucz_atlas12_2005} models with $T=4500$~K and $T=5450$~K, respectively, while faculae are represented using the FALP model \citep{fontenla_calculation_1999} as modified by \citet{unruh_spectral_1999}.
Since the secular variability analysed here is dominated by the treatment of the quiet-Sun component discussed below, these substitutions have only a minor effect on the resulting TSI reconstructions.

Unlike the active-component fluxes, the QS flux in CHRONOS varies with time.
\citet{egorova_revised_2018} defined the flux of the quiet Sun as:
\begin{equation}
F_{\mathrm{QS}}\left(\lambda,t\right)=F_C(\lambda)+\left(F_{\mathrm{min}}(\lambda)-F_C(\lambda)\right)\frac{\phi_{\mathrm{ref}}-\phi(t)}{\phi_{\mathrm{ref}}-\phi_{\mathrm{min}}},
\label{eq:qsinitial}
\end{equation}
where $F_C(\lambda)$ is the present day quiet Sun SSI \citep[here represented by the \citealt{kurucz_new_1993,kurucz_atlas12_2005} model at T = 5777 K, but \citealt{egorova_revised_2018} used model C by][]{fontenla_calculation_1999}, 
$\phi$ is the modulation potential, with the indices ``ref'' and ``min'' referring to the mean value over a reference period \citep[defined as 1974--1996;][]{egorova_revised_2018} and the minimum value of the modulation potential, which \citet[][] {egorova_revised_2018} reported to correspond roughly to 1450 during the Spörer minimum.
$F_{\mathrm{min}}(\lambda)$ represents the flux at this minimum solar activity level \citep[][] {egorova_revised_2018}.

Here, we will define the normalised variations in the modulation potential as:
\begin{equation}
    \omega(t)=\frac{\phi_{\mathrm{ref}}-\phi(t)}{\phi_{\mathrm{ref}}-\phi_{\mathrm{min}}},
    \label{eq:omega}
\end{equation}
while  the parameter:
\begin{equation}
    \gamma(\lambda)=F_{\mathrm{min}}(\lambda)-F_C(\lambda)
\end{equation}
represents the change in the QS irradiance change between the reference and the minimum-activity states. 

This allows us to rewrite  Eq. \ref{eq:qsinitial} as:
\begin{equation}
F_{\mathrm{QS}}\left(\lambda,t\right)=F_C(\lambda) + \gamma(\lambda)\omega(t).
\end{equation}
This mathematical rearrangement separates the CHRONOS reconstruction into a component independent of long-term changes in the quiet Sun and a component that explicitly captures the secular variability:
\begin{equation}
    F\left(\lambda,t\right)=F_0\left(\lambda,t\right)+ \gamma(\lambda)\omega(t)\alpha_{\mathrm{QS}}(t),
\end{equation}
where
\begin{equation}
    \label{eq:f0}
    \begin{split}
    F_0\left(\lambda,t\right) =F_{\mathrm{C}}\left(\lambda\right)&+\alpha_{\mathrm{f}}(t)(F_{\mathrm{f}}\left(\lambda\right)-F_{\mathrm{C}}\left(\lambda\right)) \\
    &+\alpha_{\mathrm{su}}(t)(F_{\mathrm{su}}\left(\lambda\right)-F_{\mathrm{C}}\left(\lambda\right)) \\
    &+\alpha_{\mathrm{sp}}(t)(F_{\mathrm{sp}}\left(\lambda\right)-F_{\mathrm{C}}\left(\lambda\right))
\end{split}
\end{equation}
and
\begin{equation}
    \alpha_{\mathrm{QS}}(t)=1-\alpha_{\mathrm{f}}(t)-\alpha_{\mathrm{su}}(t)-\alpha_{\mathrm{sp}}(t)
    \label{eq:qsff}.
\end{equation}
Integrating over wavelength gives:
\begin{equation}
    F\left(t\right)=F_0\left(t\right)+\gamma \omega(t) \alpha_{\mathrm{QS}}(t).
    \label{eq:TSI_prebelts}
\end{equation}

\subsubsection{Filling factors}
\label{sec:ff}

The filling factors describe the fractional disc coverage by different atmospheric components: faculae, sunspots, and the quiet Sun.
Following \citet{egorova_revised_2018}, we derive them empirically from ISN.
\citet{egorova_revised_2018} inferred the relationship using filling factors derived with the SATIRE-S model (where `S' refers to the satellite era) from magnetograms \citep{ball_reconstruction_2012}.

For faculae, we use \citep[Table 2 in][]{egorova_revised_2018}:
\begin{equation}
\alpha_{\mathrm{f}}(t)=(16.1\pm3)10^{-4}+(180\pm5)10^{-6}R_z(t)+(-23\pm2)10^{-8}R_z^2(t).
\label{eq:ff_f}
\end{equation}
For sunspots, we use \citep[Table 1 in][]{egorova_revised_2018}:
\begin{equation}
\alpha_{\mathrm{sp}}(t) + \alpha_{\mathrm{su}}(t) = \alpha_{\mathrm{s}}(t)=(-24\pm5)10^{-5}+(144\pm3)10^{-7}R_z(t) 
\label{eq:ff_s}
\end{equation}
Following \citet{egorova_revised_2018}, we then separate the total sunspot filling factors into umbral and penumbral by assuming a constant ratio $\alpha_{\mathrm{sp}}(t)/\alpha_{\mathrm{su}}(t)=4$ \citep[e.g.][]{wenzler_reconstruction_2006}.

\subsubsection{Activity belts}
Following \citet{krivova_reconstruction_2007,wu_solar_2018-2,egorova_revised_2018} we account for the fact that faculae and sunspots are concentrated within activity belts extending approximately to $\pm45^\circ$ and $\pm30^\circ$ latitude, respectively.
This is important because radiative properties of magnetic feature depend on their position on the solar disc.
We therefore integrate the fluxes of the active-region components only over the corresponding activity belts.
Consequently, the filling factors used in Eq.~\ref{eq:f0} and Eq.~\ref{eq:qsff} must represent the fractional coverage within these latitude ranges rather than over the full solar disc.
This is accounted by fixed scaling factors of 2.42 and 1.6 for the sunspot and facular filling factors within the $\pm30^\circ$ and $\pm45^\circ$ latitudinal belts, respectively.

Restricting the active components to activity belts changes the normalisation of their flux contrasts relative to the quiet Sun, since the active-region fluxes are no longer evaluated over the full solar disc.
To account for this, an additional correction is applied to the quiet-Sun contribution in Eq.~\ref{eq:qsff}. 
For TSI reconstructions, this results in fixed correction factors of 0.73 and 0.52 for the facular and sunspot belts, respectively, applied to the wavelength-integrated QS flux.
Equation~\ref{eq:qsff} then becomes:
\begin{equation}
\alpha_{\mathrm{QS}}(t)=(1-1.17\alpha_{\mathrm{f}}(t)-1.26\alpha_{\mathrm{su}}(t)-1.26\alpha_{\mathrm{sp}}(t)).
    \label{eq:final_tsi}
\end{equation}

\begin{table}[t!]
	\caption{Comparison between our CHRONOS reconstructions and published TSI series.}             
	\label{tab:fitresults}      
	\centering                                      
	\small
	\begin{tabular}{l*{4}{c}}  
\hline
TSI&Mod.& $\gamma$ & RMS &R\\
    &&[W~m$^{-2}$]&[W~m$^{-2}$]&\\
		\hline\hline                        
\citet{shapiro_new_2011} &SSR11 & -3.21$\pm$ 0.01 & 0.14 & 1.00\\ 
\citet{egorova_revised_2018} &SSR11 & -1.90$\pm$ 0.01 & 0.12 & 1.00\\ 
\citet{egorova_revised_2018} &MC17 & -2.82$\pm$ 0.01 & 0.19 & 0.99\\
\citet{egorova_revised_2018} &MU16 & -2.45$\pm$ 0.01 & 0.23 & 0.99\\ 
\citet{egorova_revised_2018} &US16 & -1.54$\pm$ 0.01 & 0.29 & 0.98\\
CPMDF &OW24& -0.01$\pm$ 0.04 & 0.11 & 0.97 \\
CPMDF &NMU17&-0.01$\pm$ 0.04 & 0.11 & 0.97\\
ACRIM &OW24&  0.38$\pm$ 0.11 & 0.29 & 0.85 \\
ACRIM &NMU17&0.36$\pm$ 0.11 & 0.29 & 0.85\\
Mean &OW24&  0.13$\pm$ 0.03 & 0.23 & 0.86 \\
Mean &NMU17& 0.12$\pm$ 0.03 & 0.23 & 0.86\\
\hline
\hline    
\end{tabular}
\tablefoot{The first column identifies the TSI series used both to determine $\gamma$ and to compute the corresponding metrics.
    The second column gives the modulation potential used. Listed are the fitted $\gamma$ values, together with the RMS differences and correlation coefficients ($R$) between our reconstructions and the published series. The metrics are computed over the entire period during which each pair of reconstructed and reference TSI series overlap (see Sect. \ref{sec:TSI}).} 
\footnotetext{}
\footnotetext[1]{}
\end{table}

\begin{table}[ht!]
	\caption{Comparison between direct TSI measurements and our CHRONOS reconstructions when $\gamma$ is derived based on the previous versions of CHRONOS}.            
	\label{tab:compowens}      
	\centering                                      
	\small
	\begin{tabular}{l*{6}{c}}  
\hline
Reference TSI&SSR11&\multicolumn{4}{c}{\citet{egorova_revised_2018}}\\
$\phi$ &SSR11&SSR11&MC17&MU16&US16\\
    \hline\hline       
    \multicolumn{6}{c}{RMS [W~m$^{-2}$]}\\
    \hline
CPMDF &  1.24 &  0.74 &  1.09 &  0.94 &  0.60 \\
ACRIM &  0.97 &  0.63 &  0.86 &  0.77 &  0.54\\
Mean &  1.34 &  0.85 &  1.19 &  1.05 &  0.72 \\
\hline
    \multicolumn{6}{c}{Maximum difference [W~m$^{-2}$]}\\
    \hline
CPMDF &  1.95 &  1.21 &  1.73 &  1.52 &  1.01\\
ACRIM &  1.57 &  1.18 &  1.45 &  1.34 &  1.07 \\
Mean &  2.14 &  1.49 &  1.92 &  1.74 &  1.32 \\
\hline
    \multicolumn{6}{c}{R}\\
    \hline
CPMDF &  0.49 &  0.60 &  0.51 &  0.54 &  0.65\\
ACRIM &  0.44 &  0.54 &  0.46 &  0.49 &  0.58 \\
Mean &  0.25 &  0.37 &  0.28 &  0.31 &  0.42 \\
    \hline
\end{tabular}
\tablefoot{ The first and second rows indicate the TSI records and modulation potentials used to determine $\gamma$ (see Table \ref{tab:fitresults}). The CHRONOS reconstructions use the combined \citet[][OW24]{owens_geomagnetic_2024} and \citet[][NMU17]{usoskin_heliospheric_2017} modulation potentials. 
    The metrics are computed for the entire period covered by the reference TSI series (see Sect. \ref{sec:TSI}).}
\footnotetext{}
\footnotetext[1]{}
\end{table}

\subsubsection{Secular trend}
\label{sec:lt}

The long-term variability in CHRONOS is controlled by the parameters $\omega(t)$ and $\gamma$. 
The parameter $\omega(t)$ determines the temporal evolution of the secular component through the modulation potential, while $\gamma$ defines its amplitude.
\citet{shapiro_new_2011} and \citet{egorova_revised_2018} defined the reference minimum period, min, using the lowest value of the modulation potential, corresponding to the Sp\"orer minimum around the year 1450 (being 206, 119, 376, and 11 MV for US16, MU16, MC17, and SSR11, respectively).
The quantity $F_{\mathrm{min}}(\lambda)$ therefore represents the quiet-Sun irradiance at the dimmest state of the Sun over the period covered by the modulation potentials.
Since no direct measurements exist for such a state, $F_{\mathrm{min}}(\lambda)$ must be prescribed.
The choice of $F_{\mathrm{min}}(\lambda)$ therefore depends on the adopted model atmosphere. 
\citet{shapiro_new_2011} used model~A of \citet{fontenla_calculation_1999}, which was later shown  to produce an exaggerated secular trend \citep{judge_confronting_2012}. 
\citet{egorova_revised_2018} instead adopted an atmosphere intermediate between models~A and C of  \citet{fontenla_calculation_1999}.
More recently, \citet{yeo_dimmest_2020} questioned the applicability of these atmosphere models to such a state of the Sun.

Here, instead of prescribing $F_{\mathrm{min}}(\lambda)$, we treat $\gamma$ as a free parameter and constrain it empirically by regressing Eq.~\ref{eq:TSI_prebelts} against observed TSI series. 
This approach allows the secular variability in CHRONOS to be evaluated independently of assumptions about the atmosphere structure of the dimmest quiet Sun.
Because the OW24 modulation potential does not extend back to Sp\"orer or Maunder minima, we redefine $\phi_{\mathrm{min}}$ using the period 1880--1920, during which the OW24 series reaches its lowest values.
Figure \ref{fig:modpotentials}b) shows the resulting $\omega(t)$ series for all modulation potentials considered here.

\begin{figure*}[]
   {	\centering
	\begin{overpic}[width=0.97\linewidth,trim={0 0.2cm 0cm 0.0cm},clip]{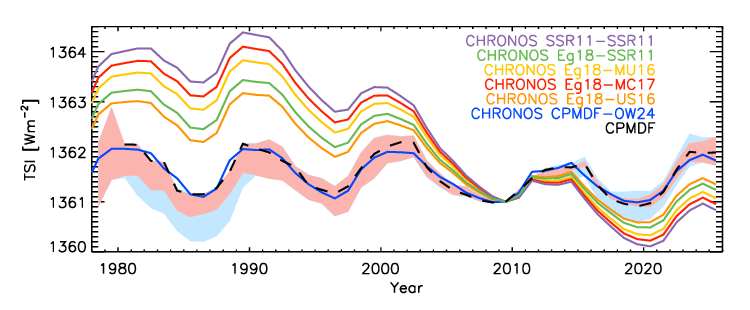}
	    \put (14.5,34) {a)}     \end{overpic}
\begin{overpic}[width=0.97\linewidth,trim={0 0.cm 0cm 0.2cm},clip]{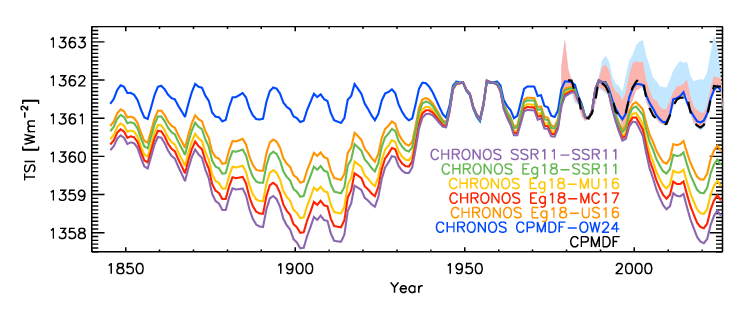}
	    \put (14.5,34) {b)}     \end{overpic}
        \begin{overpic}[width=0.97\linewidth,trim={0 0.cm 0cm 0.2cm},clip]{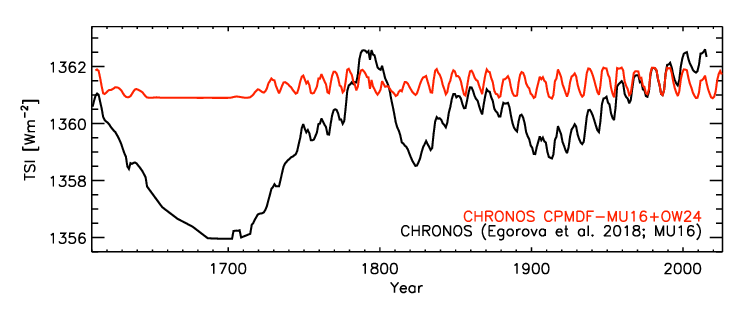}
	    \put (14.5,34) {c)}     \end{overpic}}
 	  \caption{\textit{Panels a-b: }TSI reconstructed with the CHRONOS model using the \citet{usoskin_heliospheric_2017} modulation potential from neutron monitor data connected to the OW24 series. 
      The reconstructions use $\gamma$ values obtained by fitting our CHRONOS reconstructions, based on
      the four modulation potentials employed by \citet{egorova_revised_2018}, to the corresponding TSI reconstructions of \citet[][referred to as Eg18; orange for US16, red for MC17, yellow for MU16, and green for SSR11]{egorova_revised_2018}, as well as to the \citet[][violet for SSR11]{shapiro_new_2011} reconstruction. 
      Also shown is the reconstruction obtained when $\gamma$ was constrained by regression of the OW24-based reconstruction to the \mbox{CPMDF} TSI composite (blue).
      The \mbox{CPMDF} TSI composite is shown as a dashed black curve.
      The full spread of direct TSI composites is shown as shaded regions in two cases: using only the  \mbox{CPMDF}, DdWK17, ROB, and C3S composites (pink), and including ACRIM and all 21 compilations by \citet[][cyan]{connolly_multiple_2024}. The top panel shows only the satellite era, while the bottom panel covers the full extent of the OW24 modulation potential.
      \textit{Panel c: } Comparison between the original CHRONOS TSI reconstruction by \citet{egorova_revised_2018} using the MU16 modulation potential and our reconstruction where
      $\gamma$ was determined by regression to the \mbox{CPMDF} TSI composite  and using the combined MU16+OW24 modulation potential (red).
      Shown are annual mean values.
      All series were offset to match 1361~W\,m$^{-2}$ during the 2009 (panel a) and 1986 (panels b and c) minima. }\label{fig:reconstructions}
\end{figure*}

\subsection{CHRONOS TSI reconstructions}
\label{sec:results_orchronos}

We first determine the values of $\gamma$ that allow our implementation of CHRONOS to reproduce, as closely as possible, the published TSI reconstruction of \citet{egorova_revised_2018} for each of the four modulation potentials considered in that study. 
For completeness, we perform the same analysis for the reconstruction of \citet{shapiro_new_2011}. 

Table \ref{tab:fitresults} summarises the derived $\gamma$ values together with the resulting RMS differences and linear correlation coefficients ($R$) between our reconstructions and the published CHRONOS TSI series.
Overall, we obtain very good agreement, with correlation coefficients between 0.98 and 1, indicating that our formulation reproduces the temporal behaviour and secular variability of the published reconstructions closely. 
Small differences remain between our reconstructions and the published series (see Appendix~\ref{appendix:offsets}).
They likely arise from implementation details that are not fully specified in the original publications, including possible differences in the sunspot number series used and in how different datasets were connected.
In addition, some of the original model atmospheres and auxiliary inputs used in the published CHRONOS calculations are no longer publicly available, so an exact reproduction is not possible. 
Our implementation therefore represents a close reconstruction of the published methodology rather than an exact duplication of the original calculations.

For the MC17, MU16, and US16 modulation potentials, we further found that an 11 year temporal offset was required in order to reproduce the TSI by \citet{egorova_revised_2018} and derive the corresponding $\gamma$ values.
We discuss this issue in Appendix \ref{appendix:offsets}.
Here we note only that the modulation potentials shown in Figure~\ref{fig:modpotentials} are without this offset.

The derived $\gamma$ values lie in the range [-1.54, -2.82]~W~m$^{-2}$ for the \citet{egorova_revised_2018} reconstructions and $-3.21$~W~m$^{-2}$ for the \citet{shapiro_new_2011} reconstruction.
In principle, \citet{egorova_revised_2018} used the same prescription for all modulation potentials.
The differences in $\gamma$ obtained here are primarily due to the choice of the normalisation period for the modulation potential.

Figure \ref{fig:modpotentials}c) shows the quantity $F_{\mathrm{QS}}\left(t\right)-F_C=\gamma \omega(t)$, which represents the secular component of the reconstruction relative to the reference period.
The inferred changes in the QS contribution are large: approximately $[-4.5, -5.5]$~W\,m$^{-2}$ between the Sp\"orer minimum and the reference period for the \citet{egorova_revised_2018} reconstructions and about $-8$~W\,m$^{-2}$ for \citet{shapiro_new_2011}.
These values are consistent with the secular trends of the published CHRONOS reconstructions.
For the MC17 modulation potential, $F_{\mathrm{QS}}\left(t\right)-F_C$ additionally shows an almost 2~W\,m$^{-2}$ decrease over the last three decades (see also Fig.~\ref{fig:reconstructions_offsetmc17}b). 
This behaviour was not seen in the published \citet{egorova_revised_2018} reconstructions, because it was largely suppressed by the temporal offset discussed above and in Appendix~\ref{appendix:offsets}.

Having identified the $\gamma$ values required to reproduce the published CHRONOS TSI reconstructions, we next assess whether these secular trends are consistent with modern observations.
To do so, we apply the derived $\gamma$ values to reconstruct TSI with CHRONOS using the OW24 and NMU17 modulation potentials. 
To account for uncertainties in the reconstruction procedure, we consider the full range of $\gamma$ values listed in Table~\ref{tab:fitresults}.
Table~\ref{tab:compowens} compares the resulting CHRONOS reconstructions with various measured TSI series using annual values over the entire period during which each pair of series overlaps (see Sect. \ref{sec:TSI}). 
The CHRONOS reconstructions obtained using the $\gamma$ values required to reproduce the published version of \citet{egorova_revised_2018} show relatively poor agreement with modern TSI records.
For annual values, correlations are generally below 0.7, 
RMS differences are larger than 0.57~W\,m$^{-2}$, and maximum differences exceed 1~W\,m$^{-2}$.
For comparison, the corresponding values between the SATIRE-T and \mbox{CPMDF} series are R=0.96, RMS differences 0.11~W\,m$^{-2}$, and maximum difference of 0.27 W~m$^{-2}$.
Figure~\ref{fig:reconstructions} shows that CHRONOS reconstructions with the original secular component but driven by the combined OW24 and NMU17 modulation potentials diverge substantially from direct TSI measurements.
Taken together, these results indicate that the secular variability implied by the original implementations is considerably larger than evidenced by modern TSI observations.

We therefore next determine the values of $\gamma$ that return CHRONOS reconstructions consistent with direct TSI measurements when using the OW24 modulation potential and neutron monitor data, NMU17.
Table~\ref{tab:fitresults} also lists the resulting $\gamma$ values obtained by regressing Eq.~\ref{eq:TSI_prebelts} to different TSI series, as well as the corresponding RMS and linear correlation coefficients.
Using either OW24 alone or the combined OW24+NMU17 modulation potential yields nearly identical results.
For example, comparison with the \mbox{CPMDF} composite gives $\gamma=-0.01\pm0.04$~W\,m$^{-2}$ in both cases. 
These values are about two orders of magnitude smaller than those required to reproduce the original \citet[][\ref{tab:fitresults}]{egorova_revised_2018} reconstructions.
Figure \ref{fig:reconstructions} shows that the corresponding CHRONOS reconstruction agrees closely with the \mbox{CPMDF} composite.

When using the ACRIM composite, as well as the ``Mean'' TSI series (see Sect. \ref{sec:TSI}), the regression returns positive $\gamma$ values. 
This could imply that lower solar activity corresponds to a brighter quiet Sun, opposite to the assumed physical interpretation of the CHRONOS model.
This behaviour arises because these series contain an increase across the ACRIM gap that is inconsistent with the variability inferred from  neutron monitor data.

Finally, we apply the value of $\gamma$ derived from the comparison with direct TSI measurements to the modulation potential series used by \citet{egorova_revised_2018}. 
This allows us to estimate the secular variability produced by CHRONOS when it is constrained by direct TSI measurements.
Figure~\ref{fig:reconstructions}c) compares our reconstruction based on the combined OW24+MU16 modulation potential using $\gamma$ values obtained from comparison with the \mbox{CPMDF} composite to the  \citet{egorova_revised_2018} CHRONOS reconstruction with MU16. 
Our reconstruction produces substantially weaker secular variability than the original CHRONOS series.
For example, the reconstructed TSI during the Maunder minimum is about 4.5~W\,m$^{-2}$ higher than in the original \citet{egorova_revised_2018} MU16-based reconstruction  (see Appendix~\ref{appendix:offsets} for the different behaviour over the last decades).
Table~\ref{tab:seculartrends} summarises the TSI changes between 1700 and the 1986 minimum obtained for the different CHRONOS reconstructions together with previously published estimates from \citet{egorova_revised_2018}, \citet{shapiro_new_2011}, \citet{temaj_solar_2026}, and \citet{penza_reconstruction_2024}.
The resulting increase in the observationally constrained CHRONOS TSI reconstructions between 1700 and the 1986 minimum is approximately 0.08~W\,m$^{-2}$ when using the same modulation potentials as in \citet{egorova_revised_2018}, rising to 0.1~W\,m$^{-2}$ when considering the combined OW24 and MU16 modulation potential.
These values are significantly smaller (by a factor of about 50--63) than those reported in the original \citet{egorova_revised_2018} reconstruction using the same modulation potential.
As noted above, fits based on the ACRIM and ``Mean'' composites produce positive secular trends, which are not physically consistent within the CHRONOS model.

\begin{table*}[ht!]
	\caption{Difference in TSI [W~m$^{-2}$] between the 1986 and 1700 activity minima for the PEA24 and CHRONOS reconstructions. }
	\label{tab:seculartrends}      
	\centering                                      
	\small
	\begin{tabular}{l*{8}{c}}  
\hline
&\multicolumn{2}{c}{\citet{penza_reconstruction_2024}}&\multicolumn{6}{c}{CHRONOS}\\
& OSF1&OSF2&SSR11&MU16&MC17&US16&OW24+MU16&Range\\
    \hline\hline                        
CPMDF  &0.49&-0.25& -0.08$\pm$ 0.07 & -0.08$\pm$ 0.08 & -0.08$\pm$ 0.08 & -0.09$\pm$ 0.11 & -0.10$\pm$ 0.17 & [-0.27,  0.08] \\
ACRIM  &1.30&1.30&  0.61$\pm$ 0.20 &  0.64$\pm$ 0.21 &  0.67$\pm$ 0.22 &  0.95$\pm$ 0.30 &  1.51$\pm$ 0.47 & [ 0.41,  1.98] \\
Mean &1.11&-0.20&  0.16$\pm$ 0.06 &  0.17$\pm$ 0.06 &  0.18$\pm$ 0.07 &  0.27$\pm$ 0.09 &  0.46$\pm$ 0.14 & [ 0.10,  0.60] \\
    \hline
\citet{egorova_revised_2018} &-&-& -3.51 & -5.04 & -5.46 & -4.06&-& [-5.46, -3.51]\\
\citet{shapiro_new_2011} &-&-& -5.87 & -&-&-&-&-5.87\\
\citet{penza_reconstruction_2024} &-2.12&-& -&-&-&-&- & -2.12\\
\citet{temaj_solar_2026} &-&-& -&-&-&-&- & -0.25\\
\hline
\end{tabular}
\tablefoot{The open solar flux or modulation potential series used in our reconstructions are listed in the second header row, while $N$ and $\gamma$ are determined by regression to the TSI series given in the first column. The last column gives the full range spanned by all CHRONOS reconstructions in each row.  The lower part of the table lists the corresponding secular changes for the previously published TSI reconstructions.}
\footnotetext{}
\footnotetext[1]{}
\end{table*}

\begin{table}[ht!]
	\caption{Parameters obtained by fitting our PEA24 reconstructions to composites of direct TSI measurements.}
	\label{tab:fitresultsowens}      
	\centering                                      
	\small
	\begin{tabular}{l*{16}{c}}  
\hline
    TSI&OSF& N & RMS&R \\
	&&& [Wm$^{-2}$]&\\
    \hline\hline                        
CPMDF &OSF1& -0.11$\pm$ 0.03 & 0.14 & 0.93 \\  
CPMDF &OSF2& 0.004$\pm$ 0.030 & 0.15 & 0.92\\
CPMDF 20/01/2023 &OSF1& -0.15$\pm$ 0.03 & 0.15 & 0.93 \\
CPMDF 20/01/2023 &OSF2&-0.02$\pm$ 0.03 & 0.15 & 0.92\\

ACRIM &OSF1& -0.23$\pm$ 0.04 & 0.31 & 0.81 \\
ACRIM &OSF2&-0.19$\pm$ 0.04 & 0.33 & 0.78\\
Mean & OSF1&-0.20$\pm$ 0.05 & 0.25 & 0.82 \\
Mean  &OSF2&-0.05$\pm$ 0.05 & 0.28 & 0.79\\
 \hline
\end{tabular}
\tablefoot{The TSI series used for the regression are listed in the first column, while the second header row indicates the OSF records used for the reconstruction.}
\footnotetext{}
\footnotetext[1]{}
\end{table}

\section{Reassessing the model by \citet{penza_reconstruction_2024}}
\label{sec:penza}
\subsection{Model description}

This model was introduced by \citet{penza_total_2022} and later updated by \citet{penza_reconstruction_2024}.
It is an empirical TSI reconstruction of the form:
\begin{equation}
    F(t)=F_{\mathrm{QS}}+C_n\left[1+F_{\mathrm{LT}}(t)\right]N+\alpha_{\mathrm{f}}(t) \delta_{\mathrm{fn}}+\alpha_{\mathrm{s}}(t) \delta_{\mathrm{s}},
\end{equation}
where $F_{\mathrm{QS}}$, $C_n$, $N$, $\delta_{\mathrm{fn}}$, and $\delta_{\mathrm{s}}$ are constants, while $F_{\mathrm{LT}}(t)$ (LT for long-term) describes the secular component of the reconstruction not directly associated with active-region faculae and sunspots.
Following
\citet{penza_reconstruction_2024}, we adopt: $C_n=8.8\times10^{-4}\pm1\times10^{-4}$, $\delta_{\mathrm{fn}}=0.041\pm0.002$, $\delta_{\mathrm{s}}=-0.30\pm0.01$, and $F_{\mathrm{QS}}=1359.83$~W\,m$^{-2}$.
The only parameter we allow to vary is $N$, which controls the amplitude of the secular component  (they employed $N=0.55$). 
We determine $N$ empirically by fitting the reconstruction to various TSI series.

\subsubsection{Filling factors}

Following \citet{penza_reconstruction_2024}, the sunspot and facular filling factors are based on the sunspot area record of \citet{mandal_sunspot_2020} and the plage areas of \citet{chatzistergos_analysis_2020}  (Sect. \ref{sec:sunspotnumber}).
These datasets, however, extend back only to 1874 and 1892, respectively, and the plage series becomes sufficiently complete only after about 1903. 
To cover earlier periods, \citet{penza_reconstruction_2024} derived filling factors empirically from open solar flux (OSF) series.
As our focus here is on the amplitude of the secular trend rather than on reconstructing the detailed short-term variability, we employ a simplified extension of the filling factors prior to the observational period.
In particular, we extend the sunspot filling factors back to 1610 by linearly scaling the group sunspot number series of \citet{hoyt_group_1998}. 
For monthly mean group numbers, $G_n$, we use: 
\begin{equation}
    \alpha_{\mathrm{s}}(t)=2.3\times10^{-4}  G_n.
\end{equation}
We derived the scaling factor from a linear regression between the monthly mean values of the \citet{hoyt_group_1998} group number series and the \citet{mandal_sunspot_2020} sunspot area series, considering all months for which both records overlap.
Plage filling factors prior to 1903 are then estimated from the sunspot filling factors using Eq.~5 of \citet{penza_reconstruction_2024}:
\begin{equation}
    \alpha_{\mathrm{f}}(t)=(1.3\pm0.1)\alpha_{\mathrm{s}}(t)^{(0.61\pm0.02)}+(0.0042\pm0.0005).
\end{equation}
This simplified treatment is sufficient for the present analysis because the secular variability in the PEA24 framework is controlled primarily through $F_{\mathrm{LT}}(t)$ and the scaling parameter $N$.

\subsubsection{Secular trend}
\label{sec:penzaseculartrend}

The secular trend in the PEA24 model is introduced through  $F_{\mathrm{LT}}(t)$, while its amplitude is controlled by the scaling parameter $N$.
\citet{penza_reconstruction_2024} constructed $F_{\mathrm{LT}}(t)$ from OSF1 (see Sect.~\ref{sec:modpotential}) using  empirical mode decomposition \citep[EMD;][]{huang_empirical_1998}.
The OSF1 series was decomposed into intrinsic mode functions with different characteristic timescales, and the components with timescales longer than 22 years were summed to obtain $F_{\mathrm{LT}}(t)$.

We applied the same approach to the two OSF series considered here, and the resulting long-term components are shown in Fig. \ref{fig:modpotentials}d).
However, the EMD approach becomes increasingly unstable near the edges of the series (see Appendix \ref{appendix:smoothing}), precisely where comparison with direct TSI measurements is required to fix the scaling parameter $N$.
Another drawback of EMD is mode mixing, whereby oscillations of different characteristic scales are combined within a single component \citep{wu_ensemble_2009}, making it difficult to separate and remove variability based on a given characteristic period.
To reduce the uncertainty due to these effects, we take a slightly modified approach and derive $F_{\mathrm{LT}}(t)$ by applying a 14-year low-pass filter based on SSA.
The filter width was chosen to reproduce, as closely as possible, the long-term variability obtained with the EMD approach of \citet{penza_reconstruction_2024}.
To minimise edge artefacts in the SSA approach, we ignore the first and last 14~years of the smoothed series.
The resulting SSA-based long-term components are also shown in Fig.~\ref{fig:modpotentials}d).
For comparison, we additionally applied the original EMD approach to OSF1 truncated at 2020, following \citet{penza_reconstruction_2024}.
In the following, unless otherwise stated, all OSF results presented for the PEA24 model are based on the SSA-smoothed OSF series.

\begin{figure*}[]
  {	\centering
	\begin{overpic}[width=0.96\linewidth,trim={0 0.2cm 0cm 0.0cm},clip]{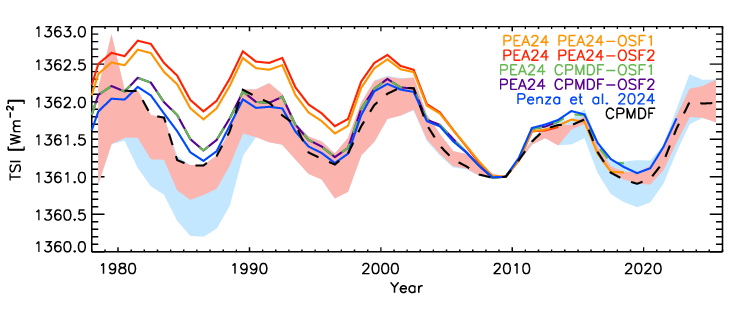}
	    \put (16.5,31) {}     \end{overpic}
\begin{overpic}[width=0.96\linewidth,trim={0 0.cm 0cm 0.2cm},clip]{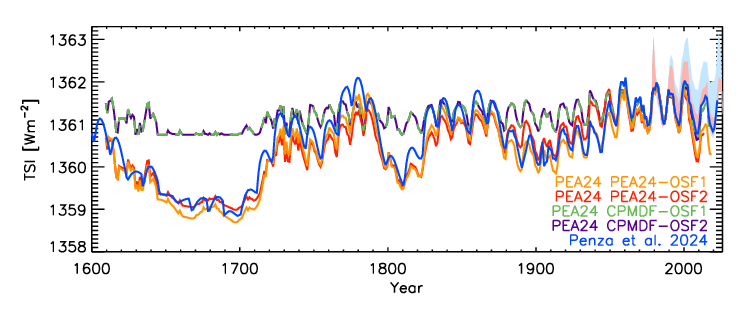}
	    \put (16.5,31) {}     \end{overpic}}
 	  \caption{TSI reconstructed with the model by \citet{penza_reconstruction_2024}. 
      Shown are the reconstructions based on OSF1 and OSF2, with the scaling parameter $N$ determined either by fitting to the original \citet{penza_reconstruction_2024} reconstruction (orange and red, respectively) or to the CPMDF TSI composite (green and purple, respectively; nearly overlapping).
    The original \citet{penza_reconstruction_2024} reconstruction is shown in blue.
    Also shown (only in the top panel) is the CPMDF TSI composite (dashed black).
      The shaded regions indicate the spread of direct TSI composite records in two cases: considering only CPMDF, DdWK17, ROB, and C3S (pink), and additionally including ACRIM and the 21 series of \citet[][cyan]{connolly_multiple_2024}. Shown are annual mean values offset to match 1361~W\,m$^{-2}$ over the 2009 minimum in the upper panel and the 1986 minimum in the lower panel. 
      The top panel shows the satellite era, while the bottom panel extends back to the Maunder minimum.
	  }\label{fig:penzareconstructions}
\end{figure*}

\subsection{TSI reconstruction with the model by \citet{penza_reconstruction_2024}}

Figure~\ref{fig:penzareconstructions} and Table \ref{tab:fitresultsowens} summarise our TSI reconstruction using the PEA24 model.
We produced separate reconstructions using OSF1 and OSF2 and determined the scaling parameter $N$ by comparison with various TSI series.

We first determined the value of $N$ required to reproduce the secular trend of the original \citet{penza_reconstruction_2024} reconstruction.
For OSF1, we obtained $N=0.31\pm0.03$, compared to $N=0.55$ reported by \citet{penza_reconstruction_2024} and $0.38\pm0.02$ adopted by \citet{penza_total_2022}. 
We also computed $N$ using the version of the CPMDF TSI composite dated 20 January 2023, which is closer to the version available to \citet{penza_reconstruction_2024}. 
The results, presented in Table~\ref{tab:fitresultsowens}, show that the derived value of $N$ is even lower than that obtained using the more recent version of the CPMDF composite, but with OSF2 the value is still consistent with 0 within the uncertainty.
To assess the impact of the smoothing method, we additionally repeated the analysis using the original EDM approach applied to OSF1 truncated at 2020, following \citet{penza_reconstruction_2024}.
Its secular variability agrees closely with the published reconstruction, although we still obtained $N=0.3\pm0.03$.
This indicates that the lower fitted values of $N$ are not primarily caused by the change from EMD to SSA smoothing.

For both OSF1 and OSF2, reconstructions using $N$ values chosen to reproduce the original \citet{penza_reconstruction_2024} secular trend, progressively diverge from direct TSI measurements. 
When aligned to the 2009 activity minimum (Fig. \ref{fig:penzareconstructions}), the discrepancy reaches approximately 0.5~W\,m$^{-2}$ during the 1986 minimum.

Comparison with direct TSI composites, in contrast, returns the best agreement for negative values of $N$ or $N\approx0$. 
In particular, only the reconstruction using OSF2 and CPMDF TSI composite as the reference yields a positive value of $N$, although it remains consistent with 0 within the uncertainty. 
Likewise, the value of $N$ derived using OSF2 and the ``Mean'' TSI composite series is consistent with 0 within the uncertainty. 
Negative values of $N$ would imply that the long-term component $F_{\mathrm{LT}}(t)$ acts in the opposite sense to that assumed in the model formulation and would therefore not be physically meaningful within the PEA24 model.
The fact that direct TSI composites favour $N\approx0$ does not imply that small-scale magnetic fields or network have no influence on long-term TSI variability in general.
Rather, it indicates that within this particular model and parametrisation adopted by \citet{penza_total_2022,penza_reconstruction_2024}, the direct TSI measurements do not require an additional long-term component introduced through $F_{\mathrm{LT}}(t)$.
This interpretation is also consistent with \citet{temaj_solar_2026}, who found that the long-term contribution of the small-scale elements is modest compared to that of active regions.

Table \ref{tab:seculartrends} summarises the TSI differences between the 1986 and 1700 minima obtained in our TSI reconstructions with the PEA24 model too.
For OSF2, fitting to the direct TSI composites, except ACRIM, yields a secular change of approximately -$0.2 - -0.25 $~W\,m$^{-2}$. 
We note that, for OSF1, as well as for OSF2 fitted to ACRIM, the reconstructions yield a TSI during the Maunder Minimum that exceeds the 1986 minimum by 0.49--1.3~W\,m$^{-2}$. 
This is an artefact of the fitting procedure returning a negative value of $N$ for these cases and is therefore unrealistic. 
The derived trends remain substantially smaller than the $-2.12$~W\,m$^{-2}$ secular change in the original \citet{penza_reconstruction_2024} reconstruction.
Overall, our analysis suggests that the secular variability in the original \citet{penza_total_2022,penza_reconstruction_2024} reconstructions is likely substantially overestimated relative to that supported by modern TSI observations.

\section{Discussion}
\label{sec:discussion}

Our analysis here relies on the improved modulation potential and open solar flux reconstructions by \citet{owens_geomagnetic_2024} and \citet{lockwood_reconstruction_2024}, respectively, to link neutron monitor and cosmogenic isotope records.
We emphasise that \citet{owens_geomagnetic_2024} demonstrated excellent agreement between their reconstruction, the neutron-monitor-based modulation potential (NMU17), and the $^{14}$C-based reconstruction of \citet{brehm_eleven-year_2021}, providing confidence that the large-scale evolution of the modulation potential is robust, although uncertainties remain.
Furthermore, our CHRONOS results are essentially unchanged when the modulation potential of \citet{owens_geomagnetic_2024} is replaced by the $^{14}$C-based one of \citet{brehm_eleven-year_2021} connected to the NMU17 one.

An independent constraint on the secular trends suggested by the reconstructions of \citet{egorova_revised_2018} and \citet{penza_total_2022} is provided by the simulations of \citet{rempel_contribution_2020}.
They showed that a change of 10~G in the mean vertical QS magnetic field produces a TSI variation of about 0.14\%.
The secular trends implied by \citet{egorova_revised_2018} and \citet{penza_total_2022} would therefore translate into about 11~G and 18--29~G changes in the mean quiet-Sun field, respectively.
Such large variations appear inconsistent with available observations of the QS magnetic field \citep[e.g.,][]{livingston_suns_2003,livingston_sun-as--star_2007,buhler_quiet_2013,lites_solar_2014,korpi-lagg_solar-cycle_2022}.
Long-term variations on centennial timescales cannot be ruled out entirely, existing measurements provide little support for changes of the magnitude required by these reconstructions.
In particular, over the period covered by modern observations of the QS magnetic field, no measurable changes in the mean QS magnetic field have been detected, whereas the modulation potential from neutron monitor data shows a clear secular decline.
This argues against using the modulation potential to ascribe hypothetical QS irradiance variations.
This is also consistent with our results, which yield values of $\gamma$ in CHRONOS and $N$ in PEA24 that are both close to zero within uncertainties.

We further note that the CHRONOS and PEA24 reconstructions derived here return lower TSI values during recent activity minima than during the Maunder Minimum. 
This likely reflects a limitation of the modelling approaches rather than a physically meaningful result.
Both models attribute long-term variations in the modulation potential and open solar flux to changes in the quiet Sun. 
However, these quantities also contain contributions from active-region evolution and long-term changes in the heliospheric magnetic field.
Consequently, interpreting long-term variations in modulation potential or open solar flux entirely as quiet-Sun irradiance variability is likely to exaggerate the contribution attributed to the quiet Sun.
The satellite-era TSI record spans only a limited range of long-term solar variability and therefore may not fully constrain possible centennial-scale changes in the quiet Sun.
The exact contribution of QS variability to secular TSI changes thus remains uncertain, although our results indicate that it is substantially smaller than implied by the original CHRONOS and PEA24 reconstructions.

\begin{figure*}[]
	\centering
\begin{overpic}[width=0.85\linewidth,trim={0 0.cm 0cm 0.cm},clip]{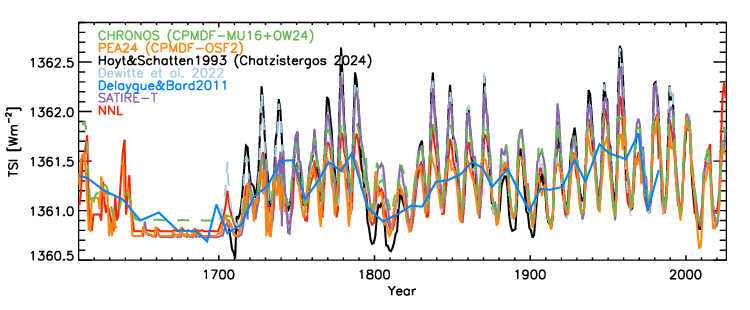}
	    \put (16.5,31) {}     \end{overpic}
 	  \caption{Comparison between the CHRONOS and PEA24 TSI reconstructions produced in this study and other long-term TSI reconstructions. 
      Our CHRONOS reconstruction was done with the MU16 modulation potential, while our PEA24 reconstruction with OSF2, the free parameters of both models were determined by regression to the CPMDF TSI composite.
      Shown are annual mean values offset to match 1361~W\,m$^{-2}$ over the 1975 minimum. However, we note that this offsetting is not optimal for the \citet{delaygue_antarctic_2011} series. We nevertheless include it to demonstrate that its relative secular trend is similarly small to those of the other series.} \label{fig:longtermreconstructions}
\end{figure*}

Overall, our analysis here, together with the recent review of existing irradiance reconstructions by \citet{chatzistergos_long-term_2023}, suggests that contemporary models increasingly favour a relatively small secular trend in irradiance, typically below 1~W\,m$^{-2}$ between the Maunder Minimum and the late twentieth century.
Nevertheless, claims of substantially larger secular variability continue to appear in the literature.
For example, \citet{connolly_how_2021, connolly_challenges_2023} argued that large secular trends remains plausible based on a sub-set of so-called ``high-variability'' TSI reconstructions.
However, most of the cited reconstructions are either outdated or rely on assumptions that have subsequently been revised. 
In particular, six of the reconstructions considered by \citet{connolly_how_2021, connolly_challenges_2023} correspond to different versions of the CHRONOS and PEA24 models.
Our analysis shows that, when updated with improved modulation potential and OSF series and evaluated against modern TSI observations, these models no longer support large secular trends.
The remaining reconstructions cited by \citet{connolly_how_2021, connolly_challenges_2023} have also been superseded by later developments.
The model of  \citet{solanki_solar_1998} evolved into the SATIRE family of models \citep[the most recent version is by][]{temaj_solar_2026}, which yield substantially lower secular variability.
Similarly, \citet{lean_reconstruction_1995} represents an early version of the Naval Research Laboratory (NRL) model, which has since undergone extensive revision \citep[most recently by][and the model was renamed to NNL]{deland_spectral_2026}. 
The reconstruction by \citet{hoyt_discussion_1993} was shown by \citet{chatzistergos_discussion_2024} to contain undocumented adjustments and inconsistencies in the input data. 
After updating and extending the reconstruction to the present, \citet{chatzistergos_discussion_2024} 
found that its secular trend was overestimated by roughly a factor of five, corresponding to a most probable value of about 0.5~W\,m$^{-2}$.
Finally, the reconstruction by \citet{bard_solar_2000} relied on a simplified linear scaling between cosmogenic isotope production rates and irradiance variations, without accounting for non-linear geomagnetic effects. 
Later revisions by \citet{delaygue_antarctic_2011}, scaling the derived modulation potential from cosmogenic isotope concentrations, indicate substantially smaller secular trends.
Taken together, current irradiance reconstructions provide little support for a large secular increase in TSI since the Maunder Minimum as shown in Fig. \ref{fig:longtermreconstructions}.

\section{Summary and conclusions}
\label{sec:summary}
Constraining the secular trend in TSI remains challenging, and published reconstructions exhibit substantial disagreement.
Estimates of the TSI increase between the 1700 and 1986 minima range from nearly zero to 6~W\,m$^{-2}$.
Although many of the high-variability reconstructions (with a TSI increase larger than 1~W\,m$^{-2}$) are now outdated and have been superseded by their newer versions, significant differences still persist among some contemporary models.

In particular, CHRONOS \citep{egorova_revised_2018} yields a TSI increase of approximately $3.5$ -- 5.5~W\,m$^{-2}$ between the 1700 and 1986 minima, while the model by \citet{penza_total_2022,penza_reconstruction_2024} yields an increase of 2.1~W\,m$^{-2}$.
Despite their different architectures, the CHRONOS and PEA24 models share a common feature: they attribute long-term irradiance variability to modulation potential or open solar flux derived from cosmogenic isotope data. 
A major uncertainty in this approach arises from the need to combine cosmogenic isotope records with neutron monitor data, which do not directly overlap.
Additional uncertainties arise from the treatment of the most recent radiocarbon data, affected by the Suess effect, as well as from the exact smoothing procedures applied to the neutron monitor data.

In this study, we re-examined the CHRONOS and PEA24 models using recent reconstructions of the modulation potential and open solar flux derived from geomagnetic indices \citep{owens_geomagnetic_2024, lockwood_reconstruction_2024}, which improve the consistency between cosmogenic-isotope records and neutron-monitor measurements, although some uncertainties remain.
Additionally, we re-evaluated both models using updated input series and a more robust smoothing methodology.
We reformulated the CHRONOS model to isolate explicitly the contribution of the quiet-Sun secular component and constrained its amplitude directly from comparisons with modern TSI observations.

Our analysis indicates that the magnitude of secular variability implied by the original CHRONOS and PEA24 reconstructions is not supported by direct TSI measurements, irrespectively of the chosen TSI composite.
When using the updated input series and constraining the secular component (via the $\gamma$ parameter in CHRONOS and $N$ in PEA24) against modern TSI measurements, the resulting reconstructions yield considerably more modest long-term changes. 
For CHRONOS, the implied TSI increase between the 1700 and 1986 minima lies in the range of roughly 0.1--0.44~W\,m$^{-2}$, depending on the chosen TSI composite, while for PEA24 the corresponding change is around 0.20 -- 0.25~W\,m$^{-2}$.
The revised CHRONOS and PEA24 reconstructions derived here are therefore broadly consistent with other contemporary TSI reconstructions, such as SATIRE-T \citep{temaj_solar_2026} and NNL \citep{deland_spectral_2026} (Fig. \ref{fig:longtermreconstructions}).

Overall, our results suggest that irradiance variations arising from hypothesised QS changes are substantially smaller (if any) than assumed in the original CHRONOS and PEA24 reconstructions. 
This conclusion is further supported by measurements of the QS magnetic field and corresponding irradiance estimates from 3D MHD simulations \citep[][see Sec. \ref{sec:discussion}]{rempel_contribution_2020}.
Furthermore, we argue that modulation potential and open solar flux are not good proxies for QS irradiance variability. 
Both quantities contain contributions from active-region evolution and long-term changes in the heliospheric magnetic field, so using them directly to parametrise QS irradiance variability is likely to overestimate its contribution.
Nevertheless, we cannot completely rule out variability associated with small-scale magnetic fields on timescales longer than those covered by current observations, though the modelling of \citet{temaj_solar_2026} suggests that any such contribution is smaller than that of active regions.
Although uncertainties in centennial-scale TSI variability remain, our results suggest that current irradiance models are increasingly converging towards a relatively modest secular increase in TSI between the 1700 and 1986 minima, likely below about 1~W\,m$^{-2}$.
Improving proxy records, such as cosmogenic isotopes, sunspots, and \ca, as well as modelling approaches, will be essential for further constraining the range of long-term TSI variability.

\section*{Data Availability}
The PEA24 reconstruction based on OSF2 and regressed to the CPMDF TSI composite, as well as the CHRONOS reconstructions obtained using all four modulation potential series (SSR11, US16, MU16, and MC17) extended with OW24 and NMU17, are available through the CDS.

\begin{acknowledgements}
We thank Matthias Rempel for insightful and stimulating discussions, that strengthened the conclusions of this work.
We also thank the anonymous reviewer who helped improve this manuscript.
This project has received funding from the European Research Council (ERC) under the European Union's Horizon 2020 research and innovation programme (grant agreement No. 101097844 — project WINSUN).
The work of TE was supported by the Swiss Karbacher Fond, Graubunden, and the Swiss National Science Foundation (SNSF) projects AEON (grant No.200020E_219166) and STOA (grant no. 200021L-22814 10001350).
MJO is part-funded by Science and Technology Facilities Council (STFC) grant number UKRI1207 and Natural Environment Research Council (NERC) grant number NE/Y001052/1.
This study has made use of SAO/NASA Astrophysics Data System's bibliographic services.
\end{acknowledgements}

\bibliographystyle{aa}
\bibliography{biblio01}

@article{frohlich_solar_2006,
	title = {Solar {Irradiance} {Variability} {Since} 1978.  {Revision} of the {PMOD} {Composite} during {Solar} {Cycle} 21},
	volume = {125},
	issn = {0038-6308},
	url = {http://adsabs.harvard.edu/abs/2006SSRv..125...53F},
	doi = {10.1007/s11214-006-9046-5},
	urldate = {2013-10-21},
	journal = {Space Science Reviews},
	author = {Fröhlich, C.},
	month = aug,
	year = {2006},
	pages = {53--65},
}

@article{ermolli_recent_2013,
	title = {Recent variability of the solar spectral irradiance and its impact on climate modelling},
	volume = {13},
	issn = {1680-7316},
	url = {http://adsabs.harvard.edu/abs/2013ACP....13.3945E},
	doi = {10.5194/acp-13-3945-2013;},
	urldate = {2013-12-06},
	journal = {Atmospheric Chemistry \& Physics},
	author = {Ermolli, I. and Matthes, K. and Dudok de Wit, T. and Krivova, N. A. and Tourpali, K. and Weber, M. and Unruh, Y. C. and Gray, L. and Langematz, U. and Pilewskie, P. and Rozanov, E. and Schmutz, W. and Shapiro, A. and Solanki, S. K. and Woods, T. N.},
	month = apr,
	year = {2013},
	pages = {3945--3977},
}

@article{unruh_spectral_1999,
	title = {The spectral dependence of facular contrast and solar irradiance variations},
	volume = {345},
	issn = {0004-6361},
	url = {http://adsabs.harvard.edu/abs/1999A%26A...345..635U},
	urldate = {2014-04-08},
	journal = {Astronomy and Astrophysics},
	author = {Unruh, Y. C. and Solanki, S. K. and Fligge, M.},
	month = may,
	year = {1999},
	pages = {635--642},
}

@article{shapiro_new_2011,
	title = {A new approach to the long-term reconstruction of the solar irradiance leads to large historical solar forcing},
	volume = {529},
	issn = {0004-6361},
	url = {http://adsabs.harvard.edu/abs/2011A&A...529A..67S},
	doi = {10.1051/0004-6361/201016173;},
	urldate = {2013-11-25},
	journal = {Astronomy and Astrophysics},
	author = {Shapiro, A. I. and Schmutz, W. and Rozanov, E. and Schoell, M. and Haberreiter, M. and Shapiro, A. V. and Nyeki, S.},
	month = may,
	year = {2011},
	pages = {67},
}

@article{gray_solar_2010,
	title = {Solar {Influences} on {Climate}},
	volume = {48},
	issn = {8755-1209},
	url = {http://adsabs.harvard.edu/abs/2010RvGeo..48.4001G},
	doi = {10.1029/2009RG000282},
	urldate = {2013-11-25},
	journal = {Reviews of Geophysics},
	author = {Gray, L. J. and Beer, J. and Geller, M. and Haigh, J. D. and Lockwood, M. and Matthes, K. and Cubasch, U. and Fleitmann, D. and Harrison, G. and Hood, L. and Luterbacher, J. and Meehl, G. A. and Shindell, D. and van Geel, B. and White, W.},
	month = oct,
	year = {2010},
	pages = {4001},
}

@article{haigh_sun_2007,
	title = {The {Sun} and the {Earth}'s {Climate}},
	volume = {4},
	url = {http://adsabs.harvard.edu/abs/2007LRSP....4....2H},
	doi = {10.12942/lrsp-2007-2},
	urldate = {2013-11-25},
	journal = {Living Reviews in Solar Physics},
	author = {Haigh, Joanna D.},
	month = oct,
	year = {2007},
	pages = {2},
}

@article{wenzler_reconstruction_2006,
	title = {Reconstruction of solar irradiance variations in cycles 21-23 based on surface magnetic fields},
	volume = {460},
	issn = {0004-6361},
	url = {http://adsabs.harvard.edu/abs/2006A%26A...460..583W},
	doi = {10.1051/0004-6361:20065752},
	urldate = {2013-11-25},
	journal = {Astronomy and Astrophysics},
	author = {Wenzler, T. and Solanki, S. K. and Krivova, N. A. and Fröhlich, C.},
	month = dec,
	year = {2006},
	pages = {583--595},
}

@article{krivova_reconstruction_2007,
	title = {Reconstruction of solar total irradiance since 1700 from the surface magnetic flux},
	volume = {467},
	issn = {0004-6361},
	url = {http://adsabs.harvard.edu/abs/2007A&A...467..335K},
	doi = {10.1051/0004-6361:20066725},
	urldate = {2014-04-08},
	journal = {Astronomy and Astrophysics},
	author = {Krivova, N. A. and Balmaceda, L. and Solanki, S. K.},
	month = may,
	year = {2007},
	pages = {335--346},
}

@article{krivova_reconstruction_2003,
	title = {Reconstruction of solar irradiance variations in cycle 23:  {Is} solar surface magnetism the cause?},
	volume = {399},
	issn = {0004-6361},
	shorttitle = {Reconstruction of solar irradiance variations in cycle 23},
	url = {http://adsabs.harvard.edu/abs/2003A&A...399L...1K},
	doi = {10.1051/0004-6361:20030029},
	urldate = {2013-10-21},
	journal = {Astronomy and Astrophysics},
	author = {Krivova, N. A. and Solanki, S. K. and Fligge, M. and Unruh, Y. C.},
	month = feb,
	year = {2003},
	pages = {L1--L4},
}

@article{fontenla_calculation_1999,
	title = {Calculation of {Solar} {Irradiances}. {I}. {Synthesis} of the {Solar} {Spectrum}},
	volume = {518},
	issn = {0004-637X},
	url = {http://adsabs.harvard.edu/abs/1999ApJ...518..480F},
	doi = {10.1086/307258},
	urldate = {2013-11-18},
	journal = {The Astrophysical Journal},
	author = {Fontenla, Juan and White, Oran R. and Fox, Peter A. and Avrett, Eugene H. and Kurucz, Robert L.},
	month = jun,
	year = {1999},
	pages = {480--499},
}

@article{solanki_solar_2013-1,
	title = {Solar {Irradiance} {Variability} and {Climate}},
	volume = {51},
	issn = {0066-4146},
	url = {http://www.annualreviews.org/doi/abs/10.1146/annurev-astro-082812-141007},
	doi = {10.1146/annurev-astro-082812-141007},
	number = {1},
	urldate = {2013-09-30},
	journal = {Annual Review of Astronomy and Astrophysics},
	author = {Solanki, Sami K. and Krivova, Natalie A. and Haigh, Joanna D.},
	month = aug,
	year = {2013},
	pages = {311--351},
}

@article{asvestari_empirical_2016,
	title = {An empirical model of heliospheric cosmic ray modulation on long-term time scale},
	volume = {6},
	url = {http://dx.doi.org/10.1051/swsc/2016011},
	doi = {10.1051/swsc/2016011},
	journal = {J. Space Weather Space Clim.},
	author = {Asvestari, Eleanna and Usoskin, Ilya G.},
	year = {2016},
	pages = {A15},
}

@article{delaygue_antarctic_2011,
	title = {An {Antarctic} view of {Beryllium}-10 and solar activity for the past millennium},
	volume = {36},
	issn = {0930-7575},
	url = {http://adsabs.harvard.edu/abs/2011ClDy...36.2201D},
	doi = {10.1007/s00382-010-0795-1},
	urldate = {2015-11-26},
	journal = {Climate Dynamics},
	author = {Delaygue, Gilles and Bard, Edouard},
	month = jun,
	year = {2011},
	pages = {2201--2218},
}

@article{lean_reconstruction_1995,
	title = {Reconstruction of solar irradiance since 1610: {Implications} for climate change},
	volume = {22},
	issn = {0094-8276},
	shorttitle = {Reconstruction of solar irradiance since 1610},
	url = {http://adsabs.harvard.edu/abs/1995GeoRL..22.3195L},
	doi = {10.1029/95GL03093},
	urldate = {2015-11-26},
	journal = {Geophysical Research Letters},
	author = {Lean, Judith and Beer, Juerg and Bradley, Raymond},
	year = {1995},
	pages = {3195--3198},
}

@article{scherrer_helioseismic_2012,
	title = {The {Helioseismic} and {Magnetic} {Imager} ({HMI}) {Investigation} for the {Solar} {Dynamics} {Observatory} ({SDO})},
	volume = {275},
	issn = {0038-0938},
	url = {http://adsabs.harvard.edu/abs/2012SoPh..275..207S},
	doi = {10.1007/s11207-011-9834-2},
	urldate = {2016-01-03},
	journal = {Solar Physics},
	author = {Scherrer, P. H. and Schou, J. and Bush, R. I. and Kosovichev, A. G. and Bogart, R. S. and Hoeksema, J. T. and Liu, Y. and Duvall, T. L. and Zhao, J. and Title, A. M. and Schrijver, C. J. and Tarbell, T. D. and Tomczyk, S.},
	month = jan,
	year = {2012},
	pages = {207--227},
}

@article{livingston_suns_2003,
	title = {The {Sun}'s immutable basal quiet atmosphere},
	volume = {212},
	issn = {0038-0938},
	url = {http://adsabs.harvard.edu/abs/2003SoPh..212..227L},
	doi = {10.1023/A:1022994002653},
	urldate = {2016-01-23},
	journal = {Solar Physics},
	author = {Livingston, W. and Wallace, L.},
	month = feb,
	year = {2003},
	pages = {227--237},
}

@article{livingston_sun-as--star_2007,
	title = {Sun-as-a-{Star} {Spectrum} {Variations} 1974-2006},
	volume = {657},
	issn = {0004-637X},
	url = {http://adsabs.harvard.edu/abs/2007ApJ...657.1137L},
	doi = {10.1086/511127},
	urldate = {2016-01-23},
	journal = {The Astrophysical Journal},
	author = {Livingston, W. and Wallace, L. and White, O. R. and Giampapa, M. S.},
	month = mar,
	year = {2007},
	pages = {1137--1149},
}

@article{ball_reconstruction_2012,
	title = {Reconstruction of total solar irradiance 1974-2009},
	volume = {541},
	issn = {0004-6361},
	url = {http://adsabs.harvard.edu/abs/2012A\%26A...541A..27B},
	doi = {10.1051/0004-6361/201118702},
	urldate = {2016-04-25},
	journal = {Astronomy and Astrophysics},
	author = {Ball, W. T. and Unruh, Y. C. and Krivova, N. A. and Solanki, S. and Wenzler, T. and Mortlock, D. J. and Jaffe, A. H.},
	month = may,
	year = {2012},
	pages = {A27},
}

@article{hoyt_group_1998,
	title = {Group {Sunspot} {Numbers}: {A} {New} {Solar} {Activity} {Reconstruction}},
	volume = {179},
	issn = {0038-0938},
	shorttitle = {Group {Sunspot} {Numbers}},
	url = {http://adsabs.harvard.edu/abs/1998SoPh..179..189H},
	doi = {10.1023/A:1005007527816},
	urldate = {2016-05-11},
	journal = {Solar Physics},
	author = {Hoyt, Douglas V. and Schatten, Kenneth H.},
	year = {1998},
	pages = {189--219},
}

@article{lockwood_centennial_2014,
	title = {Centennial variations in sunspot number, open solar flux, and streamer belt width: 1. {Correction} of the sunspot number record since 1874},
	volume = {119},
	issn = {0148-0227},
	shorttitle = {Centennial variations in sunspot number, open solar flux, and streamer belt width},
	url = {http://adsabs.harvard.edu/abs/2014JGRA..119.5172L},
	doi = {10.1002/2014JA019970},
	urldate = {2016-06-10},
	journal = {Journal of Geophysical Research (Space Physics)},
	author = {Lockwood, M. and Owens, M. J. and Barnard, L.},
	month = jul,
	year = {2014},
	pages = {5172--5182},
}

@article{domingo_solar_2009,
	title = {Solar {Surface} {Magnetism} and {Irradiance} on {Time} {Scales} from {Days} to the 11-{Year} {Cycle}},
	volume = {145},
	issn = {0038-6308, 1572-9672},
	url = {http://link.springer.com/article/10.1007/s11214-009-9562-1},
	doi = {10.1007/s11214-009-9562-1},
	language = {en},
	number = {3-4},
	urldate = {2016-07-06},
	journal = {Space Science Reviews},
	author = {Domingo, V. and Ermolli, I. and Fox, P. and Fröhlich, C. and Haberreiter, M. and Krivova, N. and Kopp, G. and Schmutz, W. and Solanki, S. K. and Spruit, H. C. and Unruh, Y. and Vögler, A.},
	month = jul,
	year = {2009},
	pages = {337--380},
}

@article{buhler_quiet_2013,
	title = {Quiet {Sun} magnetic fields observed by {Hinode}: {Support} for a local dynamo},
	volume = {555},
	issn = {0004-6361},
	shorttitle = {Quiet {Sun} magnetic fields observed by {Hinode}},
	url = {http://adsabs.harvard.edu/abs/2013A\%26A...555A..33B},
	doi = {10.1051/0004-6361/201321152},
	urldate = {2016-09-01},
	journal = {Astronomy and Astrophysics},
	author = {Bühler, D. and Lagg, A. and Solanki, S. K.},
	month = jul,
	year = {2013},
	pages = {A33},
}

@article{lites_solar_2014,
	title = {The solar cycle dependence of the weak internetwork flux},
	volume = {66},
	issn = {0004-6264},
	url = {http://adsabs.harvard.edu/abs/2014PASJ...66S...4L},
	doi = {10.1093/pasj/psu082},
	urldate = {2016-09-01},
	journal = {Publications of the Astronomical Society of Japan},
	author = {Lites, Bruce W. and Centeno, Rebecca and McIntosh, Scott W.},
	month = dec,
	year = {2014},
	pages = {S4},
}

@article{muscheler_revised_2016,
	title = {The {Revised} {Sunspot} {Record} in {Comparison} to {Cosmogenic} {Radionuclide}-{Based} {Solar} {Activity} {Reconstructions}},
	volume = {291},
	issn = {0038-0938, 1573-093X},
	url = {http://link.springer.com/article/10.1007/s11207-016-0969-z},
	doi = {10.1007/s11207-016-0969-z},
	language = {en},
	number = {9-10},
	urldate = {2016-11-24},
	journal = {Solar Physics},
	author = {Muscheler, Raimund and Adolphi, Florian and Herbst, Konstantin and Nilsson, Andreas},
	month = nov,
	year = {2016},
	note = {Number: 9-10},
	pages = {3025--3043},
}

@article{willson_total_1997,
	title = {Total solar irradiance trend during solar cycles 21 and 22.},
	volume = {277},
	issn = {0036-8075},
	url = {http://adsabs.harvard.edu/abs/1997Sci...277.1963W},
	doi = {10.1126/science.277.5334.1963},
	urldate = {2017-01-15},
	journal = {Science},
	author = {Willson, R. C.},
	month = sep,
	year = {1997},
	pages = {1963--1965},
}

@article{willson_composite_2003,
	title = {Composite total solar irradiance time series show a secular 0.04 \%/decade trend},
	volume = {31},
	url = {http://adsabs.harvard.edu/abs/2003AGUFMSH31C..06W},
	urldate = {2017-01-15},
	journal = {AGU Fall Meeting Abstracts},
	author = {Willson, R. C. and Mordvinov, A. V.},
	month = dec,
	year = {2003},
}

@article{solanki_solar_1998,
	title = {Solar irradiance since 1874 revisited},
	volume = {25},
	url = {https://ui.adsabs.harvard.edu/\#abs/1998GeoRL..25..341S/abstract},
	doi = {10.1029/98GL50038},
	urldate = {2017-01-17},
	journal = {Geophysical Research Letters},
	author = {Solanki, S. K. and Fligge, M.},
	year = {1998},
}

@article{hoyt_discussion_1993,
	title = {A discussion of plausible solar irradiance variations, 1700-1992},
	volume = {98},
	issn = {0148-0227},
	url = {http://adsabs.harvard.edu/abs/1993JGR....9818895H},
	doi = {10.1029/93JA01944},
	urldate = {2017-01-17},
	journal = {Journal of Geophysical Research},
	author = {Hoyt, Douglas V. and Schatten, Kenneth H.},
	month = nov,
	year = {1993},
	pages = {18},
}

@book{beer_cosmogenic_2012,
	edition = {1},
	series = {Physics of {Earth} and {Space} {Environments}},
	title = {Cosmogenic {Radionuclides} - {Theory} and {Applications} in the {Terrestrial} and {Space} {Environments}},
	isbn = {978-3-642-14651-0},
	url = {http://www.springer.com/us/book/9783642146503},
	language = {English},
	urldate = {2017-01-23},
	publisher = {Springer-Verlag Berlin Heidelberg},
	author = {Beer, Jürg and McCracken, Ken and von Steiger, Rudolf},
	year = {2012},
}

@article{kurucz_atlas12_2005,
	title = {{ATLAS12}, {SYNTHE}, {ATLAS9}, {WIDTH9}, et cetera},
	volume = {8},
	issn = {0037-8720},
	url = {http://adsabs.harvard.edu/abs/2005MSAIS...8...14K},
	urldate = {2017-03-05},
	journal = {Memorie della Societa Astronomica Italiana Supplementi},
	author = {Kurucz, Robert L.},
	year = {2005},
	pages = {14},
}

@article{usoskin_solar_2016,
	title = {Solar activity during the {Holocene}: the {Hallstatt} cycle and its consequence for grand minima and maxima},
	volume = {587},
	issn = {0004-6361},
	shorttitle = {Solar activity during the {Holocene}},
	url = {http://adsabs.harvard.edu/abs/2016A\%26A...587A.150U},
	doi = {10.1051/0004-6361/201527295},
	urldate = {2017-03-13},
	journal = {Astronomy and Astrophysics},
	author = {Usoskin, I. G. and Gallet, Y. and Lopes, F. and Kovaltsov, G. A. and Hulot, G.},
	year = {2016},
	pages = {A150},
}

@article{krivova_acrim-gap_2009,
	title = {{ACRIM}-gap and total solar irradiance revisited: {Is} there a secular trend between 1986 and 1996?},
	volume = {36},
	issn = {0094-8276},
	shorttitle = {{ACRIM}-gap and total solar irradiance revisited},
	url = {http://adsabs.harvard.edu/abs/2009GeoRL..3620101K},
	doi = {10.1029/2009GL040707},
	urldate = {2017-03-18},
	journal = {Geophysical Research Letters},
	author = {Krivova, N. A. and Solanki, S. K. and Wenzler, T.},
	month = oct,
	year = {2009},
	pages = {L20101},
}

@article{mccracken_geomagnetic_2004,
	title = {Geomagnetic and atmospheric effects upon the cosmogenic {10Be} observed in polar ice},
	volume = {109},
	issn = {0148-0227},
	url = {http://adsabs.harvard.edu/abs/2004JGRA..109.4101M},
	doi = {10.1029/2003JA010060},
	urldate = {2017-03-20},
	journal = {Journal of Geophysical Research (Space Physics)},
	author = {McCracken, K. G.},
	month = apr,
	year = {2004},
	pages = {A04101},
}

@article{usoskin_solar_2005,
	title = {Solar activity, cosmic rays, and {Earth}'s temperature: {A} millennium-scale comparison},
	volume = {110},
	issn = {0148-0227},
	shorttitle = {Solar activity, cosmic rays, and {Earth}'s temperature},
	url = {http://adsabs.harvard.edu/abs/2005JGRA..11010102U},
	doi = {10.1029/2004JA010946},
	urldate = {2017-03-20},
	journal = {Journal of Geophysical Research (Space Physics)},
	author = {Usoskin, I. G. and Schüssler, M. and Solanki, S. K. and Mursula, K.},
	month = oct,
	year = {2005},
	pages = {A10102},
}

@article{vonmoos_large_2006,
	title = {Large variations in {Holocene} solar activity: {Constraints} from {10Be} in the {Greenland} {Ice} {Core} {Project} ice core},
	volume = {111},
	issn = {0148-0227},
	shorttitle = {Large variations in {Holocene} solar activity},
	url = {http://adsabs.harvard.edu/abs/2006JGRA..11110105V},
	doi = {10.1029/2005JA011500},
	urldate = {2017-03-20},
	journal = {Journal of Geophysical Research (Space Physics)},
	author = {Vonmoos, Maura and Beer, Jürg and Muscheler, Raimund},
	month = oct,
	year = {2006},
	pages = {A10105},
}

@article{suess_radiocarbon_1955,
	title = {Radiocarbon {Concentration} in {Modern} {Wood}},
	volume = {122},
	copyright = {Copyright © 1955 by the American Association for the Advancement of Science},
	issn = {0036-8075, 1095-9203},
	url = {http://science.sciencemag.org/content/122/3166/415.2},
	doi = {10.1126/science.122.3166.415-a},
	language = {en},
	number = {3166},
	urldate = {2017-04-08},
	journal = {Science},
	author = {Suess, Hans E.},
	month = sep,
	year = {1955},
	note = {Number: 3166},
	pages = {415--417},
}

@article{muscheler_solar_2007,
	title = {Solar activity during the last 1000 yr inferred from radionuclide records},
	volume = {26},
	issn = {0277-3791},
	url = {http://adsabs.harvard.edu/abs/2007QSRv...26...82M},
	doi = {10.1016/j.quascirev.2006.07.012},
	urldate = {2017-05-03},
	journal = {Quaternary Science Reviews},
	author = {Muscheler, Raimund and Joos, Fortunat and Beer, Jürg and Müller, Simon A. and Vonmoos, Maura and Snowball, Ian},
	month = jan,
	year = {2007},
	pages = {82--97},
}

@article{judge_confronting_2012,
	title = {Confronting a solar irradiance reconstruction with solar and stellar data},
	volume = {544},
	issn = {0004-6361},
	url = {http://adsabs.harvard.edu/abs/2012A&A...544A..88J},
	doi = {10.1051/0004-6361/201218903},
	urldate = {2017-05-25},
	journal = {Astronomy and Astrophysics},
	author = {Judge, P. G. and Lockwood, G. W. and Radick, R. R. and Henry, G. W. and Shapiro, A. I. and Schmutz, W. and Lindsey, C.},
	month = aug,
	year = {2012},
	pages = {A88},
}

@article{chatzistergos_new_2017,
	title = {New reconstruction of the sunspot group numbers since 1739 using direct calibration and “backbone” methods},
	volume = {602},
	copyright = {© ESO, 2017},
	issn = {0004-6361, 1432-0746},
	url = {https://doi.org/10.1051/0004-6361/201630045},
	doi = {10.1051/0004-6361/201630045},
	language = {en},
	urldate = {2017-06-15},
	journal = {Astronomy \& Astrophysics},
	author = {Chatzistergos, Theodosios and Usoskin, Ilya G. and Kovaltsov, Gennady A. and Krivova, Natalie A. and Solanki, Sami K.},
	month = jun,
	year = {2017},
	pages = {A69},
}

@article{shapiro_nature_2017,
	title = {The nature of solar brightness variations},
	volume = {1},
	issn = {2397-3366},
	url = {http://adsabs.harvard.edu/abs/2017NatAs...1..612S},
	doi = {10.1038/s41550-017-0217-y},
	journal = {Nature Astronomy},
	author = {Shapiro, A. I. and Solanki, S. K. and Krivova, N. A. and Cameron, R. H. and Yeo, K. L. and Schmutz, W. K.},
	month = sep,
	year = {2017},
	pages = {612--616},
}

@article{yeo_solar_2017,
	title = {Solar {Irradiance} {Variability} is {Caused} by the {Magnetic} {Activity} on the {Solar} {Surface}},
	volume = {119},
	issn = {0031-9007},
	url = {http://adsabs.harvard.edu/abs/2017PhRvL.119i1102Y},
	doi = {10.1103/PhysRevLett.119.091102},
	journal = {Physical Review Letters},
	author = {Yeo, K. L. and Solanki, S. K. and Norris, C. M. and Beeck, B. and Unruh, Y. C. and Krivova, N. A.},
	month = sep,
	year = {2017},
}

@article{egorova_revised_2018,
	title = {Revised historical solar irradiance forcing},
	volume = {615},
	issn = {0004-6361},
	url = {http://adsabs.harvard.edu/abs/2018A&A...615A..85E},
	doi = {10.1051/0004-6361/201731199},
	urldate = {2018-09-04},
	journal = {Astronomy and Astrophysics},
	author = {Egorova, T. and Schmutz, W. and Rozanov, E. and Shapiro, A. I. and Usoskin, I. and Beer, J. and Tagirov, R. V. and Peter, T.},
	month = jul,
	year = {2018},
	pages = {A85},
}

@article{wu_solar_2018-2,
	title = {Solar total and spectral irradiance reconstruction over the last 9000 years},
	volume = {620},
	copyright = {© ESO 2018},
	issn = {0004-6361, 1432-0746},
	url = {https://www.aanda.org/articles/aa/full_html/2018/12/aa32956-18/aa32956-18.html},
	doi = {10.1051/0004-6361/201832956},
	language = {en},
	urldate = {2018-12-08},
	journal = {Astronomy \& Astrophysics},
	author = {Wu, C.-J. and Krivova, N. A. and Solanki, S. K. and Usoskin, I. G.},
	month = dec,
	year = {2018},
	pages = {A120},
}

@article{lean_estimating_2018,
	title = {Estimating {Solar} {Irradiance} {Since} 850 {CE}},
	volume = {5},
	copyright = {©2018. The Authors.},
	issn = {2333-5084},
	url = {https://agupubs.onlinelibrary.wiley.com/doi/abs/10.1002/2017EA000357},
	doi = {10.1002/2017EA000357},
	language = {en},
	number = {4},
	urldate = {2019-03-06},
	journal = {Earth and Space Science},
	author = {Lean, J. L.},
	year = {2018},
	pages = {133--149},
}

@article{dewitte_total_2016,
	title = {The {Total} {Solar} {Irradiance} {Climate} {Data} {Record}},
	volume = {830},
	issn = {0004-637X},
	url = {http://adsabs.harvard.edu/abs/2016ApJ...830...25D},
	doi = {10.3847/0004-637X/830/1/25},
	urldate = {2020-01-22},
	journal = {The Astrophysical Journal},
	author = {Dewitte, Steven and Nevens, Stijn},
	month = oct,
	year = {2016},
	pages = {25},
}

@article{chatzistergos_analysis_2020,
	title = {Analysis of full-disc {Ca} {II} {K} spectroheliograms - {III}. {Plage} area composite series covering 1892–2019},
	volume = {639},
	copyright = {© T. Chatzistergos et al. 2020},
	issn = {0004-6361, 1432-0746},
	url = {https://www.aanda.org/articles/aa/full_html/2020/07/aa37746-20/aa37746-20.html},
	doi = {10.1051/0004-6361/202037746},
	language = {en},
	urldate = {2020-07-13},
	journal = {Astronomy \& Astrophysics},
	author = {Chatzistergos, Theodosios and Ermolli, Ilaria and Krivova, Natalie A. and Solanki, Sami K. and Banerjee, Dipankar and Barata, Teresa and Belik, Marcel and Gafeira, Ricardo and Garcia, Adriana and Hanaoka, Yoichiro and Hegde, Manjunath and Klimeš, Jan and Korokhin, Viktor V. and Lourenço, Ana and Malherbe, Jean-Marie and Marchenko, Gennady P. and Peixinho, Nuno and Sakurai, Takashi and Tlatov, Andrey G.},
	month = jul,
	year = {2020},
	pages = {A88},
}

@article{kren_where_2017,
	title = {Where does {Earth}'s atmosphere get its energy?},
	volume = {7},
	url = {http://adsabs.harvard.edu/abs/2017JSWSC...7A..10K},
	doi = {10.1051/swsc/2017007},
	urldate = {2020-07-14},
	journal = {Journal of Space Weather and Space Climate},
	author = {Kren, Andrew C. and Pilewskie, Peter and Coddington, Odele},
	month = mar,
	year = {2017},
	pages = {A10},
}

@inproceedings{kurucz_new_1993,
	address = {San Francisco},
	series = {International {Astronomical} {Union}, {Colloquium} {No}. 138, held in {Trieste}, {Italy}, {July} 1992},
	title = {A {New} {Opacity}-{Sampling} {Model} {Atmosphere} {Program} for {Arbitrary} {Abundances}},
	volume = {44},
	isbn = {0-937707-63-5},
	url = {http://adsabs.harvard.edu/abs/1993ASPC...44...87K},
	urldate = {2020-07-14},
	booktitle = {Astronomical {Society} of the {Pacific}},
	publisher = {Astronomical Society of the Pacific},
	author = {Kurucz, R. L.},
	editor = {dworetsky, M and Castelli, F and Faraggiana, R},
	month = jan,
	year = {1993},
	pages = {87},
}

@article{mandal_sunspot_2020,
	title = {Sunspot area catalog revisited: {Daily} cross-calibrated areas since 1874},
	volume = {640},
	copyright = {© S. Mandal et al. 2020},
	issn = {0004-6361, 1432-0746},
	shorttitle = {Sunspot area catalog revisited},
	url = {https://www.aanda.org/articles/aa/abs/2020/08/aa37547-20/aa37547-20.html},
	doi = {10.1051/0004-6361/202037547},
	language = {en},
	urldate = {2020-09-01},
	journal = {Astronomy \& Astrophysics},
	author = {Mandal, Sudip and Krivova, Natalie A. and Solanki, Sami K. and Sinha, Nimesh and Banerjee, Dipankar},
	month = aug,
	year = {2020},
	pages = {A78},
}

@article{dudok_de_wit_methodology_2017,
	title = {Methodology to create a new total solar irradiance record: {Making} a composite out of multiple data records},
	volume = {44},
	issn = {0094-8276},
	shorttitle = {Methodology to create a new total solar irradiance record},
	url = {https://agupubs.onlinelibrary.wiley.com/doi/full/10.1002/2016GL071866},
	doi = {10.1002/2016GL071866},
	number = {3},
	urldate = {2020-09-09},
	journal = {Geophysical Research Letters},
	author = {Dudok de Wit, Thierry and Kopp, Greg and Fröhlich, Claus and Schöll, Micha},
	month = feb,
	year = {2017},
	pages = {1196--1203},
}

@article{yeo_dimmest_2020,
	title = {The {Dimmest} {State} of the {Sun}},
	volume = {47},
	copyright = {©2020. The Authors.},
	issn = {1944-8007},
	url = {https://agupubs.onlinelibrary.wiley.com/doi/abs/10.1029/2020GL090243},
	doi = {10.1029/2020GL090243},
	language = {en},
	number = {19},
	urldate = {2020-10-12},
	journal = {Geophysical Research Letters},
	author = {Yeo, K. L. and Solanki, S. K. and Krivova, N. A. and Rempel, M. and Anusha, L. S. and Shapiro, A. I. and Tagirov, R. V. and Witzke, V.},
	year = {2020},
	pages = {e2020GL090243},
}

@article{bard_solar_2000,
	title = {Solar irradiance during the last 1200 years based on cosmogenic nuclides},
	volume = {52},
	issn = {0280-6509},
	url = {http://adsabs.harvard.edu/abs/2000TellB..52..985B},
	doi = {10.3402/tellusb.v52i3.17080},
	urldate = {2020-12-15},
	journal = {Tellus Series B Chemical and Physical Meteorology B},
	author = {Bard, Edouard and Raisbeck, Grant and Yiou, Françoise and Jouzel, Jean},
	month = jul,
	year = {2000},
	pages = {985--992},
}

@article{brehm_eleven-year_2021,
	title = {Eleven-year solar cycles over the last millennium revealed by radiocarbon in tree rings},
	copyright = {2021 The Author(s), under exclusive licence to Springer Nature Limited},
	issn = {1752-0908},
	url = {https://www.nature.com/articles/s41561-020-00674-0},
	doi = {10.1038/s41561-020-00674-0},
	language = {en},
	urldate = {2021-01-05},
	journal = {Nature Geoscience},
	author = {Brehm, Nicolas and Bayliss, Alex and Christl, Marcus and Synal, Hans-Arno and Adolphi, Florian and Beer, Jürg and Kromer, Bernd and Muscheler, Raimund and Solanki, Sami K. and Usoskin, Ilya and Bleicher, Niels and Bollhalder, Silvia and Tyers, Cathy and Wacker, Lukas},
	month = jan,
	year = {2021},
	pages = {1--6},
}

@article{rempel_contribution_2020,
	title = {On the {Contribution} of {Quiet}-{Sun} {Magnetism} to {Solar} {Irradiance} {Variations}: {Constraints} on {Quiet}-{Sun} {Variability} and {Grand}-minimum {Scenarios}},
	volume = {894},
	issn = {0004-637X},
	shorttitle = {On the {Contribution} of {Quiet}-{Sun} {Magnetism} to {Solar} {Irradiance} {Variations}},
	url = {http://adsabs.harvard.edu/abs/2020ApJ...894..140R},
	doi = {10.3847/1538-4357/ab8633},
	urldate = {2021-05-18},
	journal = {The Astrophysical Journal},
	author = {Rempel, M.},
	month = may,
	year = {2020},
	pages = {140},
}

@article{usoskin_solar_2021,
	title = {Solar cyclic activity over the last millennium reconstructed from annual {14C} data},
	volume = {649},
	copyright = {© ESO 2021},
	issn = {0004-6361, 1432-0746},
	url = {https://www.aanda.org/articles/aa/abs/2021/05/aa40711-21/aa40711-21.html},
	doi = {10.1051/0004-6361/202140711},
	language = {en},
	urldate = {2021-07-13},
	journal = {Astronomy \& Astrophysics},
	author = {Usoskin, I. G. and Solanki, S. K. and Krivova, N. A. and Hofer, B. and Kovaltsov, G. A. and Wacker, L. and Brehm, N. and Kromer, B.},
	month = may,
	year = {2021},
	pages = {A141},
}

@article{connolly_how_2021,
	title = {How much has the {Sun} influenced {Northern} {Hemisphere} temperature trends? {An} ongoing debate},
	volume = {21},
	issn = {1674-4527},
	shorttitle = {How much has the {Sun} influenced {Northern} {Hemisphere} temperature trends?},
	url = {https://iopscience.iop.org/article/10.1088/1674-4527/21/6/131},
	doi = {10.1088/1674-4527/21/6/131},
	language = {en},
	number = {6},
	urldate = {2021-08-17},
	journal = {Research in Astronomy and Astrophysics},
	author = {Connolly, Ronan and Soon, Willie and Connolly, Michael and Baliunas, Sallie and Berglund, Johan and Butler, C. John and Cionco, Rodolfo Gustavo and Elias, Ana G. and Fedorov, Valery M. and Harde, Hermann and Henry, Gregory W. and Hoyt, Douglas V. and Humlum, Ole and Legates, David R. and Lüning, Sebastian and Scafetta, Nicola and Solheim, Jan-Erik and Szarka, László and Loon, Harry van and Velasco Herrera, Víctor M. and Willson, Richard C. and Yan, Hong and Zhang, Weijia},
	month = aug,
	year = {2021},
	pages = {131},
}

@article{montillet_data_2022,
	title = {Data {Fusion} of {Total} {Solar} {Irradiance} {Composite} {Time} {Series} {Using} 41 {Years} of {Satellite} {Measurements}},
	volume = {127},
	issn = {2169-8996},
	url = {https://onlinelibrary.wiley.com/doi/abs/10.1029/2021JD036146},
	doi = {10.1029/2021JD036146},
	language = {en},
	number = {13},
	urldate = {2022-07-11},
	journal = {Journal of Geophysical Research: Atmospheres},
	author = {Montillet, J.-P. and Finsterle, W. and Kermarrec, G. and Sikonja, R. and Haberreiter, M. and Schmutz, W. and Dudok de Wit, T.},
	year = {2022},
	pages = {e2021JD036146},
}

@article{dewitte_centennial_2022,
	title = {Centennial {Total} {Solar} {Irradiance} {Variation}},
	volume = {14},
	copyright = {http://creativecommons.org/licenses/by/3.0/},
	issn = {2072-4292},
	url = {https://www.mdpi.com/2072-4292/14/5/1072},
	doi = {10.3390/rs14051072},
	language = {en},
	number = {5},
	urldate = {2022-08-24},
	journal = {Remote Sensing},
	author = {Dewitte, Steven and Cornelis, Jan and Meftah, Mustapha},
	month = jan,
	year = {2022},
	pages = {1072},
}

@book{masson-delmotte_climate_2021,
	address = {Cambridge, United Kingdom and New York, NY, USA},
	title = {Climate {Change} 2021: {The} {Physical} {Science} {Basis}. {Contribution} of {Working} {Group} {I} to the {Sixth} {Assessment} {Report} of the {Intergovernmental} {Panel} on {Climate} {Change}},
	publisher = {Cambridge University Press},
	author = {{IPCC}},
	editor = {Masson-Delmotte, V. and Zhai, P. and Pirani, A. and Connors, S.L. and Péan, C. and Berger, S. and Caud, N. and Chen, Y. and Goldfarb, L. and Gomis, M.I. and Huang, M. and Leitzell, K. and Lonnoy, E. and Matthews, J.B.R. and Maycock, T.K. and Waterfield, T. and Yelekçi, O. and Yu, R. and Zhou, B.},
	year = {2021},
	doi = {10.1017/9781009157896},
}

@article{penza_total_2022,
	title = {Total {Solar} {Irradiance} during the {Last} {Five} {Centuries}},
	volume = {937},
	issn = {0004-637X},
	url = {https://doi.org/10.3847/1538-4357/ac8a4b},
	doi = {10.3847/1538-4357/ac8a4b},
	language = {en},
	number = {2},
	urldate = {2022-10-01},
	journal = {The Astrophysical Journal},
	author = {Penza, Valentina and Berrilli, Francesco and Bertello, Luca and Cantoresi, Matteo and Criscuoli, Serena and Giobbi, Piermarco},
	month = sep,
	year = {2022},
	pages = {84},
}

@article{chatzistergos_full-disc_2022,
	title = {Full-disc {Ca} {II} {K} observations - {A} window to past solar magnetism},
	volume = {9},
	issn = {2296-987X},
	url = {https://www.frontiersin.org/articles/10.3389/fspas.2022.1038949},
	doi = {10.3389/fspas.2022.1038949},
	urldate = {2022-11-17},
	journal = {Frontiers in Astronomy and Space Sciences},
	author = {Chatzistergos, Theodosios and Krivova, Natalie A. and Ermolli, Ilaria},
	year = {2022},
}

@article{clette_recalibration_2023,
	title = {Recalibration of the {Sunspot}-{Number}: {Status} {Report}},
	volume = {298},
	issn = {1573-093X},
	shorttitle = {Recalibration of the {Sunspot}-{Number}},
	url = {https://doi.org/10.1007/s11207-023-02136-3},
	doi = {10.1007/s11207-023-02136-3},
	language = {en},
	number = {3},
	urldate = {2023-03-20},
	journal = {Solar Physics},
	author = {Clette, F. and Lefèvre, L. and Chatzistergos, T. and Hayakawa, H. and Carrasco, V. M. S. and Arlt, R. and Cliver, E. W. and Dudok de Wit, T. and Friedli, T. K. and Karachik, N. and Kopp, G. and Lockwood, M. and Mathieu, S. and Muñoz-Jaramillo, A. and Owens, M. and Pesnell, D. and Pevtsov, A. and Svalgaard, L. and Usoskin, I. G. and van Driel-Gesztelyi, L. and Vaquero, J. M.},
	month = mar,
	year = {2023},
	pages = {44},
}

@article{usoskin_history_2023,
	title = {A history of solar activity over millennia},
	volume = {20},
	issn = {1614-4961},
	url = {https://doi.org/10.1007/s41116-023-00036-z},
	doi = {10.1007/s41116-023-00036-z},
	language = {en},
	number = {1},
	urldate = {2023-05-08},
	journal = {Living Reviews in Solar Physics},
	author = {Usoskin, Ilya G.},
	month = may,
	year = {2023},
	pages = {2},
}

@article{chatzistergos_long-term_2023,
	title = {Long-term changes in solar activity and irradiance},
	volume = {252},
	issn = {1364-6826},
	url = {https://www.sciencedirect.com/science/article/pii/S1364682623001487},
	doi = {10.1016/j.jastp.2023.106150},
	urldate = {2023-10-19},
	journal = {Journal of Atmospheric and Solar-Terrestrial Physics},
	author = {Chatzistergos, Theodosios and Krivova, Natalie A. and Yeo, Kok Leng},
	month = nov,
	year = {2023},
	pages = {106150},
}

@article{connolly_challenges_2023,
	title = {Challenges in the {Detection} and {Attribution} of {Northern} {Hemisphere} {Surface} {Temperature} {Trends} {Since} 1850},
	volume = {23},
	issn = {1674-4527},
	url = {https://dx.doi.org/10.1088/1674-4527/acf18e},
	doi = {10.1088/1674-4527/acf18e},
	language = {en},
	number = {10},
	urldate = {2023-10-19},
	journal = {Research in Astronomy and Astrophysics},
	author = {Connolly, Ronan and Soon, Willie and Connolly, Michael and Baliunas, Sallie and Berglund, Johan and Butler, C. J. and Cionco, Rodolfo Gustavo and Elias, Ana G. and Fedorov, Valery M. and Harde, Hermann and Henry, Gregory W. and Hoyt, Douglas V. and Humlum, Ole and Legates, David R. and Scafetta, Nicola and Solheim, Jan-Erik and Szarka, László and Herrera, Víctor M. Velasco and Yan, Hong and Zhang, Weijia},
	month = sep,
	year = {2023},
	pages = {105015},
}

@article{usoskin_heliospheric_2005,
	title = {Heliospheric modulation of cosmic rays: {Monthly} reconstruction for 1951-2004},
	volume = {110},
	issn = {0148-0227},
	shorttitle = {Heliospheric modulation of cosmic rays},
	url = {https://ui.adsabs.harvard.edu/abs/2005JGRA..11012108U},
	doi = {10.1029/2005JA011250},
	urldate = {2023-10-29},
	journal = {Journal of Geophysical Research (Space Physics)},
	author = {Usoskin, Ilya G. and Alanko-Huotari, Katja and Kovaltsov, Gennady A. and Mursula, Kalevi},
	month = dec,
	year = {2005},
	pages = {A12108},
}

@article{chatzistergos_understanding_2024,
	title = {Understanding the secular variability of solar irradiance: the potential of {Ca} {II} {K} observations},
	volume = {14},
	copyright = {© T. Chatzistergos et al., Published by EDP Sciences 2024},
	issn = {2115-7251},
	shorttitle = {Understanding the secular variability of solar irradiance},
	url = {https://www.swsc-journal.org/articles/swsc/abs/2024/01/swsc230049/swsc230049.html},
	doi = {10.1051/swsc/2024006},
	language = {en},
	urldate = {2024-04-12},
	journal = {Journal of Space Weather and Space Climate},
	author = {Chatzistergos, Theodosios and Krivova, Natalie A. and Ermolli, Ilaria},
	year = {2024},
	pages = {9},
}

@article{chatzistergos_discussion_2024,
	title = {A {Discussion} of {Implausible} {Total} {Solar}-{Irradiance} {Variations} {Since} 1700},
	volume = {299},
	issn = {1573-093X},
	url = {https://doi.org/10.1007/s11207-024-02262-6},
	doi = {10.1007/s11207-024-02262-6},
	language = {en},
	number = {2},
	urldate = {2024-05-31},
	journal = {Solar Physics},
	author = {Chatzistergos, Theodosios},
	month = feb,
	year = {2024},
	pages = {21},
}

@article{owens_geomagnetic_2024,
	title = {A {Geomagnetic} {Estimate} of {Heliospheric} {Modulation} {Potential} over the {Last} 175 {Years}},
	volume = {299},
	issn = {1573-093X},
	url = {https://doi.org/10.1007/s11207-024-02316-9},
	doi = {10.1007/s11207-024-02316-9},
	language = {en},
	number = {6},
	urldate = {2024-06-19},
	journal = {Solar Physics},
	author = {Owens, Mathew J. and Barnard, Luke A. and Muscheler, Raimund and Herbst, Konstantin and Lockwood, Mike and Usoskin, Ilya and Asvestari, Eleanna},
	month = jun,
	year = {2024},
	pages = {84},
}

@article{connolly_multiple_2024,
	title = {Multiple {New} or {Updated} {Satellite} {Total} {Solar} {Irradiance} ({TSI}) {Composites} (1978–2023)},
	volume = {975},
	issn = {0004-637X},
	url = {https://dx.doi.org/10.3847/1538-4357/ad7794},
	doi = {10.3847/1538-4357/ad7794},
	language = {en},
	number = {1},
	urldate = {2024-10-29},
	journal = {The Astrophysical Journal},
	author = {Connolly, Ronan and Soon, Willie and Connolly, Michael and Cionco, Rodolfo Gustavo and Elias, Ana G. and Henry, Gregory W. and Scafetta, Nicola and Herrera, Víctor M. Velasco},
	month = oct,
	year = {2024},
	pages = {102},
}

@article{penza_reconstruction_2024,
	title = {Reconstruction of the {Total} {Solar} {Irradiance} {During} the {Last} {Millennium}},
	volume = {976},
	issn = {0004-637X, 1538-4357},
	url = {https://iopscience.iop.org/article/10.3847/1538-4357/ad7c49},
	doi = {10.3847/1538-4357/ad7c49},
	language = {en},
	number = {1},
	urldate = {2024-12-11},
	journal = {The Astrophysical Journal},
	author = {Penza, Valentina and Bertello, Luca and Cantoresi, Matteo and Criscuoli, Serena and Lucaferri, Lorenza and Reda, Raffaele and Ulzega, Simone and Berrilli, Francesco},
	month = nov,
	year = {2024},
	pages = {11},
}

@article{amdur_negative_2025,
	title = {Negative trend in total solar irradiance over the satellite era},
	volume = {122},
	url = {https://www.pnas.org/doi/abs/10.1073/pnas.2417155122},
	doi = {10.1073/pnas.2417155122},
	number = {11},
	urldate = {2025-03-11},
	journal = {Proceedings of the National Academy of Sciences},
	author = {Amdur, Ted and Huybers, Peter},
	month = mar,
	year = {2025},
	pages = {e2417155122},
}

@article{chatzistergos_revisiting_2025,
	title = {Revisiting the {SATIRE}-{S} irradiance reconstruction: {Heritage} of {Mt} {Wilson} magnetograms and {Ca} {II} {K} observations},
	volume = {696},
	copyright = {© The Authors 2025},
	issn = {0004-6361, 1432-0746},
	shorttitle = {Revisiting the {SATIRE}-{S} irradiance reconstruction},
	url = {https://www.aanda.org/articles/aa/abs/2025/04/aa54044-25/aa54044-25.html},
	doi = {10.1051/0004-6361/202554044},
	language = {en},
	urldate = {2025-04-18},
	journal = {Astronomy \& Astrophysics},
	author = {Chatzistergos, Theodosios and Krivova, Natalie A. and Solanki, Sami K. and Yeo, Kok Leng},
	month = apr,
	year = {2025},
	pages = {A204},
}

@article{usoskin_heliospheric_2017,
	title = {Heliospheric modulation of cosmic rays during the neutron monitor era: {Calibration} using {PAMELA} data for 2006–2010},
	volume = {122},
	copyright = {©2017. American Geophysical Union. All Rights Reserved.},
	issn = {2169-9402},
	shorttitle = {Heliospheric modulation of cosmic rays during the neutron monitor era},
	url = {https://onlinelibrary.wiley.com/doi/abs/10.1002/2016JA023819},
	doi = {10.1002/2016JA023819},
	language = {en},
	number = {4},
	urldate = {2025-05-19},
	journal = {Journal of Geophysical Research: Space Physics},
	author = {Usoskin, Ilya G. and Gil, Agnieszka and Kovaltsov, Gennady A. and Mishev, Alexander L. and Mikhailov, Vladimir V.},
	year = {2017},
	pages = {3875--3887},
}

@article{chatzistergos_assessment_2025,
	title = {Assessment of sunspot number cross-calibration approaches},
	volume = {699},
	copyright = {© The Authors 2025},
	issn = {0004-6361, 1432-0746},
	url = {https://www.aanda.org/articles/aa/abs/2025/07/aa54896-25/aa54896-25.html},
	doi = {10.1051/0004-6361/202554896},
	language = {en},
	urldate = {2025-07-06},
	journal = {Astronomy \& Astrophysics},
	author = {Chatzistergos, Theodosios and Krivova, Natalie A. and Sundermann, Hannah and Usoskin, Ilya G.},
	month = jul,
	year = {2025},
	pages = {A157},
}

@article{kopp_solar_2025,
	title = {Solar irradiance measurements},
	volume = {22},
	issn = {1614-4961},
	url = {https://doi.org/10.1007/s41116-025-00040-5},
	doi = {10.1007/s41116-025-00040-5},
	language = {en},
	number = {1},
	urldate = {2025-07-12},
	journal = {Living Reviews in Solar Physics},
	author = {Kopp, Greg},
	month = jul,
	year = {2025},
	pages = {1},
}

@article{baroni_persistent_2019,
	title = {Persistent {Draining} of the {Stratospheric} {10Be} {Reservoir} {After} the {Samalas} {Volcanic} {Eruption} (1257 {CE})},
	volume = {124},
	issn = {0148-0227},
	url = {https://ui.adsabs.harvard.edu/abs/2019JGRD..124.7082B},
	doi = {10.1029/2018JD029823},
	urldate = {2025-08-04},
	journal = {Journal of Geophysical Research (Atmospheres)},
	author = {Baroni, Mélanie and Bard, Edouard and Petit, Jean-Robert and Viseur, Sophie},
	month = jul,
	year = {2019},
	pages = {7082--7097},
}

@article{lockwood_reconstruction_2024,
	title = {Reconstruction of {Carrington} {Rotation} {Means} of {Open} {Solar} {Flux} over the {Past} 154 {Years}},
	volume = {299},
	issn = {0038-0938},
	url = {https://ui.adsabs.harvard.edu/abs/2024SoPh..299...28L},
	doi = {10.1007/s11207-024-02268-0},
	urldate = {2025-12-09},
	journal = {Solar Physics},
	author = {Lockwood, Mike and Owens, Mat},
	month = mar,
	year = {2024},
	pages = {28},
}

@article{belov_large_2000,
	title = {Large {Scale} {Modulation}: {View} {From} the {Earth}},
	volume = {93},
	issn = {1572-9672},
	shorttitle = {Large {Scale} {Modulation}},
	url = {https://doi.org/10.1023/A:1026584109817},
	doi = {10.1023/A:1026584109817},
	language = {en},
	number = {1},
	urldate = {2026-01-23},
	journal = {Space Science Reviews},
	author = {Belov, Anatoly},
	month = jul,
	year = {2000},
	pages = {79--105},
}

@article{korpi-lagg_solar-cycle_2022,
	title = {Solar-cycle variation of quiet-{Sun} magnetism and surface gravity oscillation mode},
	volume = {665},
	issn = {0004-6361},
	url = {https://ui.adsabs.harvard.edu/abs/2022A&A...665A.141K},
	doi = {10.1051/0004-6361/202243979},
	urldate = {2026-03-17},
	journal = {Astronomy and Astrophysics},
	author = {Korpi-Lagg, M. J. and Korpi-Lagg, A. and Olspert, N. and Truong, H.-L.},
	month = sep,
	year = {2022},
	pages = {A141},
}

@article{deland_spectral_2026,
	title = {Spectral {Irradiance} {Observations} and {Projections} for {Solar} {Cycle} 25},
	volume = {13},
	copyright = {© 2026. The Author(s).},
	issn = {2333-5084},
	url = {https://onlinelibrary.wiley.com/doi/abs/10.1029/2025EA004810},
	doi = {10.1029/2025EA004810},
	language = {en},
	number = {4},
	urldate = {2026-03-31},
	journal = {Earth and Space Science},
	author = {DeLand, Matthew T. and Marchenko, Sergey V. and Lean, Judith L. and Coddington, Odele},
	year = {2026},
	pages = {e2025EA004810},
}

@article{huang_empirical_1998,
	title = {The empirical mode decomposition and the {Hilbert} spectrum for nonlinear and non-stationary time series analysis},
	volume = {454},
	issn = {0080-46301364-5021},
	url = {https://ui.adsabs.harvard.edu/abs/1998RSPSA.454..903H},
	doi = {10.1098/rspa.1998.0193},
	urldate = {2026-04-08},
	journal = {Proceedings of the Royal Society of London Series A},
	author = {Huang, N. E. and Shen, Z. and Long, S. R. and Wu, M. C. and Shih, H. H. and Zheng, Q. and Yen, N.-C. and Tung, C. C. and Liu, H. H.},
	month = mar,
	year = {1998},
	pages = {903--998},
}

@article{temaj_solar_2026,
	title = {Solar irradiance reconstruction over the telescopic era using a revised photospheric magnetic field model},
	volume = {708},
	copyright = {© The Authors 2026},
	issn = {0004-6361, 1432-0746},
	url = {https://www.aanda.org/articles/aa/abs/2026/04/aa59041-26/aa59041-26.html},
	doi = {10.1051/0004-6361/202659041},
	language = {en},
	urldate = {2026-04-17},
	journal = {Astronomy \& Astrophysics},
	author = {Temaj, D. and Krivova, N. A. and Chatzistergos, T. and Solanki, S. K. and Hofer, B.},
	month = apr,
	year = {2026},
	pages = {A306},
}

@article{masarik_simulation_1999,
	title = {Simulation of particle fluxes and cosmogenic nuclide production in the {Earth}'s atmosphere},
	volume = {104},
	copyright = {Copyright 1999 by the American Geophysical Union.},
	issn = {2156-2202},
	url = {https://onlinelibrary.wiley.com/doi/abs/10.1029/1998JD200091},
	doi = {10.1029/1998JD200091},
	language = {en},
	number = {D10},
	urldate = {2026-04-18},
	journal = {Journal of Geophysical Research: Atmospheres},
	author = {Masarik, J. and Beer, J.},
	year = {1999},
	pages = {12099--12111},
}

@article{usoskin_solar_2011,
	title = {Solar modulation parameter for cosmic rays since 1936 reconstructed from ground-based neutron monitors and ionization chambers},
	volume = {116},
	issn = {0148-0227},
	url = {https://ui.adsabs.harvard.edu/abs/2011JGRA..116.2104U},
	doi = {10.1029/2010JA016105},
	urldate = {2026-05-28},
	journal = {Journal of Geophysical Research (Space Physics)},
	author = {Usoskin, Ilya G. and Bazilevskaya, Galina A. and Kovaltsov, Gennady A.},
	month = feb,
	year = {2011},
	pages = {A02104},
}

@article{temaj_reconstruction_2026,
	title = {Reconstruction of annual solar irradiance over the last three millennia},
	volume = {710},
	copyright = {© The Authors 2026},
	issn = {0004-6361, 1432-0746},
	url = {https://www.aanda.org/articles/aa/abs/2026/06/aa59432-26/aa59432-26.html},
	doi = {10.1051/0004-6361/202659432},
	language = {en},
	urldate = {2026-06-17},
	journal = {Astronomy \& Astrophysics},
	author = {Temaj, D. and Krivova, N. A. and Solanki, S. K. and Usoskin, I. G. and Chatzistergos, T.},
	month = jun,
	year = {2026},
	pages = {L24},
}

@article{mccracken_long-term_2007,
	title = {Long-term changes in the cosmic ray intensity at {Earth}, 1428–2005},
	volume = {112},
	copyright = {Copyright 2007 by the American Geophysical Union.},
	issn = {2156-2202},
	url = {https://onlinelibrary.wiley.com/doi/abs/10.1029/2006JA012117},
	doi = {10.1029/2006JA012117},
	language = {en},
	number = {A10},
	urldate = {2026-06-29},
	journal = {Journal of Geophysical Research: Space Physics},
	author = {McCracken, K. G. and Beer, J.},
	year = {2007},
}

@article{asvestari_neutron_2017,
	title = {Neutron {Monitors} and {Cosmogenic} {Isotopes} as {Cosmic} {Ray} {Energy}-{Integration} {Detectors}: {Effective} {Yield} {Functions}, {Effective} {Energy}, and {Its} {Dependence} on the {Local} {Interstellar} {Spectrum}},
	volume = {122},
	copyright = {©2017. American Geophysical Union. All Rights Reserved.},
	issn = {2169-9402},
	shorttitle = {Neutron {Monitors} and {Cosmogenic} {Isotopes} as {Cosmic} {Ray} {Energy}-{Integration} {Detectors}},
	url = {https://onlinelibrary.wiley.com/doi/abs/10.1002/2017JA024469},
	doi = {10.1002/2017JA024469},
	language = {en},
	number = {10},
	urldate = {2026-06-29},
	journal = {Journal of Geophysical Research: Space Physics},
	author = {Asvestari, Eleanna and Gil, Agnieszka and Kovaltsov, Gennady A. and Usoskin, Ilya G.},
	year = {2017},
	pages = {9790--9802},
}

@article{wu_ensemble_2009,
	title = {Ensemble empirical mode decomposition: a noise-assisted data analysis method},
	volume = {01},
	issn = {1793-5369},
	shorttitle = {Ensemble empirical mode decomposition},
	url = {https://www.worldscientific.com/doi/abs/10.1142/S1793536909000047},
	doi = {10.1142/S1793536909000047},
	number = {01},
	urldate = {2026-07-05},
	journal = {Advances in Adaptive Data Analysis},
	author = {Wu, Zhaohua and Huang, Norden E.},
	month = jan,
	year = {2009},
	note = {Publisher: World Scientific Publishing Co.},
	pages = {1--41},
}

\appendix
\section{Smoothing approaches}
\label{appendix:smoothing}
In our reassessment of the CHRONOS and PEA24 models we diverged from their processing by smoothing the timeseries with an SSA approach rather than the EMD used in PEA24 or using 22-year averages as done in CHRONOS.
We argue that SSA provides a more robust procedure to derive a smoothed timeseries for the needs of these irradiance models.
In particular, SSA is less sensitive to edge effects than the other two methods.

Figure \ref{fig:smoothingmodpotentials}b) shows the series obtained by computing 22-year averages of the OW24 modulation potential, combined with NMU17, with the central year of the final bin progressively shifted one year earlier.
We find significant differences between the series. The case used by \citet[][thick red curve]{egorova_revised_2018} exhibits a relatively flat evolution, whereas using more (less) data leads to a pronounced decrease (increase). 
This demonstrates that using the modulation potentials as 22-year averages to derive their scaling is highly sensitive to the choice of years included, and is therefore not reliable for the purposes of this study.

Similarly, Figure \ref{fig:smoothingmodpotentials}c) and d) show the series obtained by applying the EMD approach on the OW24 modulation potential combined with NMU17 and OSF1 where we limit the extend of the series.
We also find significant differences with this approach, indicating that the smoothing result is sensitive to the choice of years. 
See also Fig.~\ref{fig:modpotentials}d) and Fig. 4 of \citealt{penza_reconstruction_2024}, which clearly show that the EMD-smoothed series diverges from the original data over the last few years.

\begin{figure*}[]
   {	\centering
	\begin{overpic}[width=0.91\linewidth,trim={0 0.9cm 0cm 0.0cm},clip]{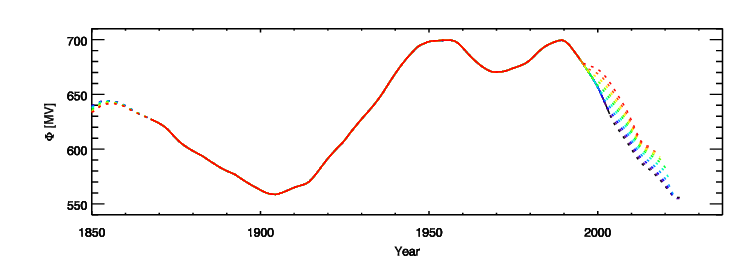}
	    \put (14.5,21) {a) SSA}     \end{overpic}
	\begin{overpic}[width=0.91\linewidth,trim={0 0.9cm 0cm 0.4cm},clip]{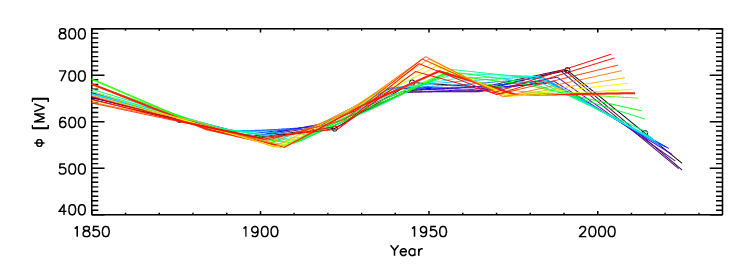}
	    \put (14.5,21) {b) 22-year averages}     \end{overpic}
	\begin{overpic}[width=0.91\linewidth,trim={0 0.9cm 0cm 0.4cm},clip]{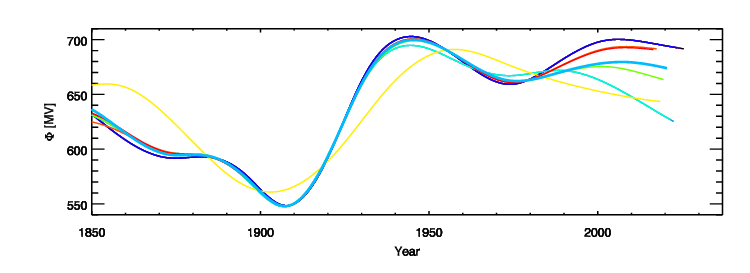}
	    \put (14.5,21) {c) EMD}     \end{overpic}
	\begin{overpic}[width=0.91\linewidth,trim={0 0.cm 0cm 0.4cm},clip]{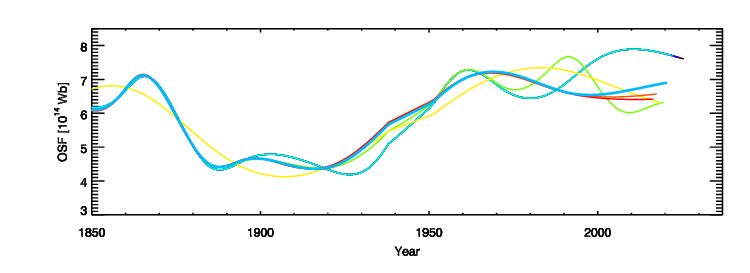} \put (14.5,28.2) {d) EMD}     \end{overpic}
 	  \caption{Effect of smoothing on the modulation potential and OSF used here.
      \textit{Panel a:} 22-year SSA smoothing of the OW24 modulation potential combined with NMU17. The black curve uses all available data, while each subsequent curve (blue, green, yellow, and red) progressively excludes one additional year from the end of the record. Solid lines indicate the central portion of the series after excluding 22 years from both ends, whereas dotted lines show the smoothing applied to the full series.
      \textit{Panel b:} 22-year averages of the  OW24 modulation potential combined with NMU17. The central year of the bins is 2014 for the black curve and shifts one year earlier for each subsequent curve (from blue, green, yellow, to red). The thick red line denotes the curve with binning as in \citet{egorova_revised_2018}.
      \textit{Panels c--d:} Application of EMD on the OW24 modulation potential combined with NMU17 and OSF1, respectively. The black curve uses all available data, while each subsequent curve (blue, green, yellow, and red) progressively excludes one additional year from the end of the record. The thick cyan curve is the ones using the same period as \citet{penza_reconstruction_2024}.
    }}
      \label{fig:smoothingmodpotentials}
\end{figure*}

\begin{figure*}[]
   {	\centering
	\begin{overpic}[width=0.91\linewidth,trim={0 0.cm 0cm 0.0cm},clip]{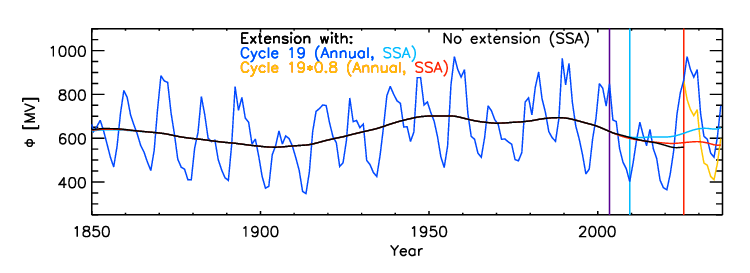}
	    \put (14.5,27) {a)}     \end{overpic}
	\begin{overpic}[width=0.91\linewidth,trim={0 0.cm 0cm 0.4cm},clip]{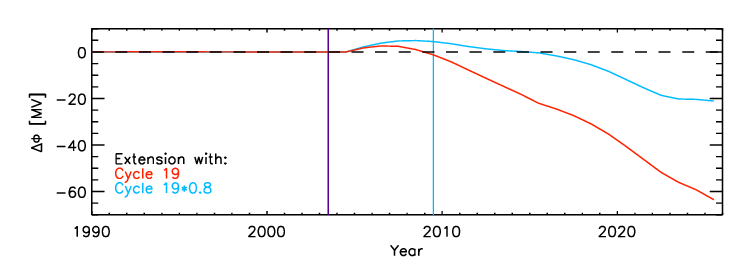} \put (14.5,25.2) {b)}     \end{overpic}
 	  \caption{\textit{Panel a:} The modulation potential of OW24 combined with NMU17, is extended by appending a repetition of cycle 19 (1957--1967) at the end of the record (blue for annual and cyan for SSA), as well as a version scaled by a factor of 0.8 (yellow for annual and red for SSA). Also shown is the original series smoothed with SSA (black). 
      \textit{Panel b:} Difference between the smoothed modulation potential of OW24, combined with NMU17, obtained using SSA with all available data, and cases in which the series is smoothed with SSA after it is extended by appending a repetition of cycle 19 (1957--1967; red) or a version of cycle 19 scaled by a factor of 0.8 (cyan). 
      The red vertical line marks the final year of observations, the cyan line indicates the 2009 solar cycle minimum, and the purple line denotes the limit beyond which the SSA smoothing is in general no longer reliable (i.e., 22 years from the last data point).}}
      \label{fig:smoothingmodpotentials1}
\end{figure*}

The SSA approach on the other hand provides a more consistent result.
Figure \ref{fig:smoothingmodpotentials}a) shows the series obtained by applying the SSA approach on the OW24 modulation potential, combined with NMU17, where we limit its extend.
There is a region near the edges where SSA is less reliable. 
In contrast to the other two methods, this region is well defined and corresponds to the smoothing window, here 22 years for our CHRONOS reconstruction and 14 for our PEA24 reconstruction.
In this study, {in our CHRONOS reconstruction} we retain the SSA-smoothed series by excluding only the first and last 16 years, rather than 22 years. 
Figure \ref{fig:smoothingmodpotentials1} illustrates why this choice is justified for our analysis. 
In particular, we consider two scenarios in which the modulation potential over the next 10 years follows either a repetition of cycle 19 (1957--1967; the strongest on record) or a version of this cycle scaled by a factor of 0.8.

\begin{figure*}[]
  {	\centering
	\begin{overpic}[width=0.96\linewidth,trim={0 0.94cm 0cm 0.0cm},clip]{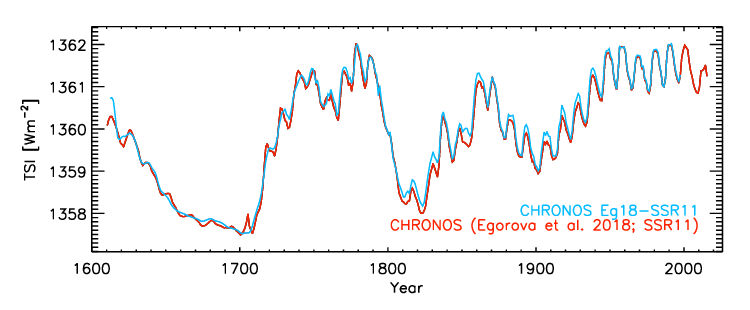}
	    \put (14.5,28) {a) SSR11}     \end{overpic}
	\begin{overpic}[width=0.96\linewidth,trim={0 0.94cm 0cm 0.34cm},clip]{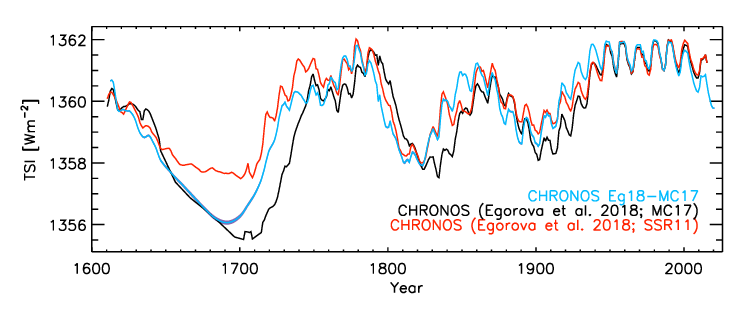}
	    \put (14.5,28) {b) MC17}     \end{overpic}
	\begin{overpic}[width=0.96\linewidth,trim={0 0.cm 0cm 0.34cm},clip]{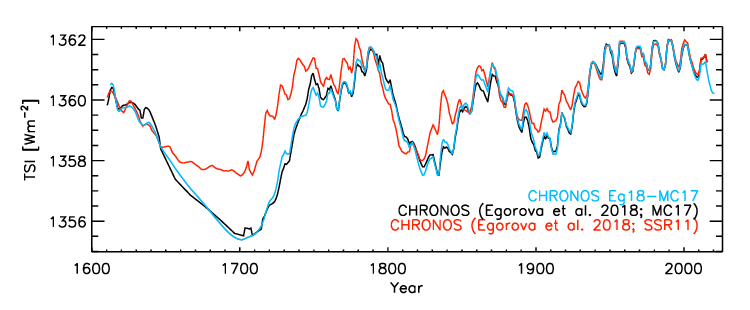}
	    \put (14.5,35) {c) MC17 (offset)}     \end{overpic}
 	  \caption{Comparison of CHRONOS TSI reconstructions with the MC17 and SSR11 modulation potentials.
     The original reconstructions of \citet{egorova_revised_2018} using SSR11 and MC17 are shown in red and black, respectively. 
     Our corresponding reconstructions are shown in cyan: panel a) uses SSR11, while panels b) and c) use MC17 without and with an 11-year offset, respectively. Shown are annual mean values which have been offset to get the value of 1361 W~m$^{-2}$ over the 1986 minimum.
	  }}\label{fig:reconstructions_offsetmc17}
\end{figure*}

\begin{figure*}[]
  {	\centering
	\begin{overpic}[width=0.96\linewidth,trim={0 0.94cm 0cm 0.34cm},clip]{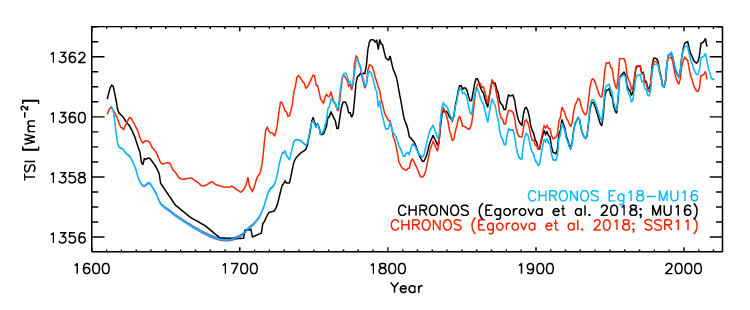}
	    \put (14.5,28) {a) MU16}     \end{overpic}
	\begin{overpic}[width=0.96\linewidth,trim={0 0.cm 0cm 0.34cm},clip]{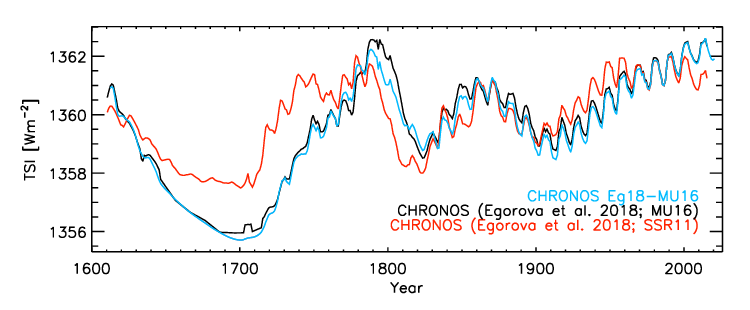}
	    \put (14.5,35) {b) MU16 (offset)}     \end{overpic}
 	  \caption{Comparison of CHRONOS TSI reconstructions using the MU16 and SSR11 modulation potentials. 
      The original reconstructions of \citet{egorova_revised_2018} based on SSR11 and MU16 are shown in red and black, respectively, while our reconstruction is shown in cyan. 
      Panels a) and b) show our reconstruction without and with an 11-year offset, respectively.
      Shown are annual mean values which have been offset to get the value of 1361 W~m$^{-2}$ over the 1986 minimum.
	  }}\label{fig:reconstructions_offsetmu16}
\end{figure*}

In the first case, we place the maximum in 2026, even though the maximum of cycle 25 has already occurred, in order to construct a conservative scenario; the more realistic case is expected to be lower, and possibly closer to the scaled case. 
We find that, in both scenarios, our SSA-smoothed series agrees closely with, or is slightly higher than, the assumed extensions between 16 and 22 years before the end of the record. 
This supports the use of SSA up to 16 years before the end of the series for our analysis, although this should be regarded as an upper bound. 
Any scenario in which cycle 25 is weaker than cycle 19 would lead to lower smoothed modulation potential values around the 2009 minimum and would therefore slightly reduce our derived scaling when regressing the reconstructed TSI to direct TSI measurements.
Thus, the CHRONOS estimate of the QS contribution to TSI would be slightly smaller than the value derived here considering that Cycle 25 is weaker than Cycle 19.

\section{Temporal offset in modulation potentials}
\label{appendix:offsets}

As we mentioned in the main text, it appears that \citet{egorova_revised_2018} shifted three of the modulation potentials by 11 years in their irradiance reconstructions.
In Figure \ref{fig:reconstructions_offsetmc17} we compare our reconstruction with the SSR11 and MC17 modulation potentials to those from \citet{egorova_revised_2018}.
We find an excellent agreement between our reconstruction and that by \citet{egorova_revised_2018} with the SSR11 modulation potential.
However, when using the MC17 in exactly the same way as the SSR11 one we get a TSI reconstruction that is clearly out of phase with the one by \citet{egorova_revised_2018}.
Figure \ref{fig:reconstructions_offsetmc17}c) shows our reconstruction with the MC17 modulation potential when we offset it by 11 years, in which case we are able to reproduce quite well the TSI series by \citet{egorova_revised_2018} with the same modulation potential.
We note that Figure \ref{fig:reconstructions_offsetmc17}b) shows that our reconstruction with MC17 without the offset exhibits a better correspondence in the timing of Maunder and Dalton minima to the \citet{egorova_revised_2018} TSI reconstruction with SSR11 than with MC17 with the temporal offset.
Surprisingly, in Fig. \ref{fig:reconstructions_offsetmc17}b) we also notice that our TSI reconstruction with MC17 exhibits a sharp decrease over cycle 24 with a decrease of about 1.5 W~m$^{-2}$ between the 1996 and 2019 minima.
Thus, the CHRONOS TSI reconstruction with MC17 modulation potential when a temporal offset is not applied would also suffice to argue against the high secular trend of the original CHRONOS reconstruction.

Figure \ref{fig:reconstructions_offsetmu16} shows the reconstructions with the MU16 modulation potential with and without the 11 year offset.
We find again that the 11 year offset is needed to reproduce the \citet{egorova_revised_2018} TSI reconstruction, as in the case with the US16 modulation potential (not shown here), but this is not the case for the SSR11 modulation potential (which we remind we digitised from Figure 2 of \citealt{shapiro_new_2011}).
Interestingly, the reconstruction with the MU16 modulation potential with the offset exhibits a rather sharp increase over the recent decades, clearly evident in the \citet{egorova_revised_2018} reconstruction with the same modulation potential, which is in stark contrast to direct measurement over that period.
This feature is removed when the offset is not used.

Both of these reconstructions hint that the secular trend in the \citet{egorova_revised_2018} TSI reconstructions was likely exaggerated.
However, in the absence of a more reliable modulation potential linking neutron monitor and cosmogenic isotope data, and given the differing behaviour of the SSR11 and US16 modulation potential, this discrepancy could not be reconciled at the time.

\end{document}